\PassOptionsToPackage{dvipsnames}{xcolor}
\documentclass[sigconf,nonacm,10pt,screen]{acmart}

\usepackage[all]{nowidow}
\usepackage{xcolor}
\usepackage{subcaption}
\usepackage[]{hyperref}
\usepackage{acronym}
\usepackage{siunitx}

\usepackage[capitalise,noabbrev]{cleveref}
\crefformat{section}{\S#2#1#3} 
\crefformat{subsection}{\S#2#1#3}
\crefformat{subsubsection}{\S#2#1#3}
\crefrangeformat{section}{\S#3#1#4 to~\S#5#2#6}
\crefrangeformat{subsection}{\S#3#1#4 to~\S#5#2#6}
\crefrangeformat{subsubsection}{\S#3#1#4 to~\S#5#2#6}

\newcommand{\trabant}{\textsf{Trabant}}

\DeclareSIUnit[]{\bps}{\text{bps}}
\DeclareSIUnit[]{\pixel}{\text{px}}

\begin{document}

\author{Tobias Pfandzelter}
\affiliation{%
    \institution{Technische Universit\"at Berlin}
    \city{Berlin}
    \country{Germany}}
\email{tp@3s.tu-berlin.de}
\orcid{0000-0002-7868-8613}

\author{Nikita Bauer}
\affiliation{%
    \institution{Technische Universit\"at Berlin}
    \city{Berlin}
    \country{Germany}}
\email{nba@3s.tu-berlin.de}
\orcid{0009-0008-1783-7575}

\author{Alexander Leis}
\affiliation{%
    \institution{Technische Universit\"at Berlin}
    \city{Berlin}
    \country{Germany}}
\email{ale@3s.tu-berlin.de}
\orcid{0009-0009-8201-2681}

\author{Corentin Perdrizet}
\affiliation{%
    \institution{Bordeaux INP ENSEIRB-MATMECA}
    \city{Bordeaux}
    \country{France}}
\email{corentin.perdrizet@bordeaux-inp.fr}
\orcid{0009-0001-7831-7169}

\author{Felix Trautwein}
\affiliation{%
    \institution{Technische Universit\"at Berlin}
    \city{Berlin}
    \country{Germany}}
\email{ftr@3s.tu-berlin.de}
\orcid{0009-0002-1573-2768}

\author{Trever Schirmer}
\affiliation{%
    \institution{Technische Universit\"at Berlin}
    \city{Berlin}
    \country{Germany}}
\orcid{0000-0001-9277-3032}
\email{ts@3s.tu-berlin.de}

\author{Osama Abboud}
\affiliation{%
    \institution{Huawei Technologies}
    \city{Munich}
    \country{Germany}}
\email{osama.abboud@huawei.com}
\orcid{0009-0003-0311-1267}

\author{David Bermbach}
\affiliation{%
    \institution{Technische Universit\"at Berlin}
    \city{Berlin}
    \country{Germany}}
\email{db@3s.tu-berlin.de}
\orcid{0000-0002-7524-3256}

\title{\trabant{}: A Serverless Architecture for Multi-Tenant Orbital Edge Computing}

\begin{abstract}
    Orbital edge computing reduces the data transmission needs of Earth observation satellites by processing sensor data on-board, allowing near-real-time insights while minimizing downlink costs.
    However, current orbital edge computing architectures are inflexible, requiring custom mission planning and high upfront development costs.
    In this paper, we propose a novel approach: shared Earth observation satellites that are operated by a central provider but used by multiple tenants.
    Each tenant can execute their own logic on-board the satellite to filter, prioritize, and analyze sensor data.

    We introduce \trabant{}, a serverless architecture for shared satellite platforms, leveraging the Func\-tion-\-as-\-a-\-Ser\-vice (FaaS) paradigm and time-shifted computing.
    This architecture abstracts operational complexities, enabling dynamic scheduling under satellite resource constraints, reducing deployment overhead, and aligning event-driven satellite observations with intermittent computation.
    We present the design of \trabant{}, demonstrate its capabilities with a proof-of-concept prototype, and evaluate it using real satellite computing telemetry data.
    Our findings suggest that \trabant{} can significantly reduce mission planning overheads, offering a scalable and efficient platform for diverse Earth observation missions.
\end{abstract}

\maketitle
\pagestyle{plain}

\section{Introduction}
\label{sec:introduction}

Orbital edge computing (OEC) can reduce the amount of data Earth observation satellites need to downlink by filtering and prioritizing sensor readings on-board the satellite, decreasing cost and energy demand while allowing insights in near-real-time~\cite{denby2020orbital,denby2023kodan,wang2023satellite,xing2024deciphering,leyva2023satellite,wang2023vision,furutanpey2024fool,yin2024comprehensive,furano2020towards}.
Current OEC architectures require careful mission planning, aligning sensor resolution, orbital parameters, energy harvesting equipment, algorithms, and compute capabilities for a specific use case, meaning OEC-equipped satellites are inflexible and have high upfront development costs~\cite{giuffrida2021varphi,denby2023kodan,wang2023satellite}.

\begin{figure}
    \centering
    \includegraphics[width=\linewidth]{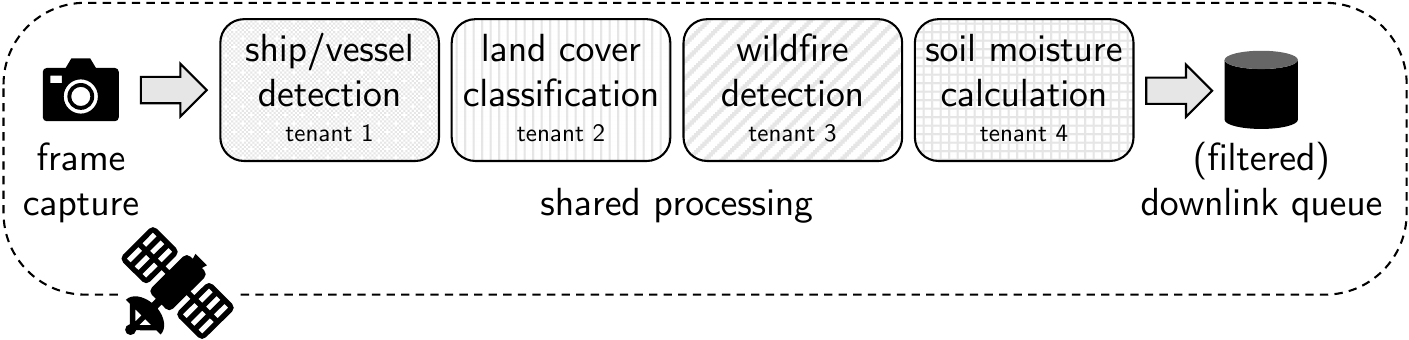}
    \caption{A shared satellite with orbital edge computing could continuously capture Earth observation data and process it using services and models from different tenants, reducing the downlink strain while still providing near-real-time insights}
    \label{fig:intro}
\end{figure}

Instead of these vertically integrated, purpose-built satellites, we propose \emph{shared} Earth observation satellites, owned and operated by a satellite operator, but simultaneously used by different tenants to flexibly execute their Earth observation missions.
We show a simplified version of this idea in \cref{fig:intro}.
Our shared satellite captures frames using its cameras and sensors, those frames and their metadata are processed on the satellite by different services, which then insert their results into a downlink queue or discard frames that are not of value to them.
Essentially, clients can use their own logic on-board the satellite to filter and prioritize Earth observation data and perform inference tasks in as soon as new data is captured.
The key insights that lead us to believe that this is a viable architecture are:

\begin{enumerate}
    \item That Earth observation satellites are often unused, e.g., a satellite meant to observe wildfires also flies over the oceans, and that sharing sensing capabilities can reduce overall mission costs.
    \item That the mission planning and operating processes between satellite operators and Earth observation clients are not well aligned: While operators need months and thousands of dollars to develop a satellite and then operate it for years to amortize those costs, clients, e.g., scientists, are interested in running new live queries, continuously updating their inference and filtering logic, or deploying new measurement campaigns, all with as little lead time as possible.
    \item That OEC already provides a flexible platform for sharing resources using the same service deployment and management approaches that are used in terrestrial computing today.
\end{enumerate}

Emphasizing this last point, we present \trabant{}, a serverless architecture for shared Earth observation satellite platforms with OEC.
We believe that following the principles of serverless computing, especially the Function-as-a-Service (FaaS) deployment paradigm, can help overcome the salient challenges of operating shared satellite platforms:

\begin{itemize}
    \item The high level of abstraction in FaaS allows satellite operators to schedule or interrupt OEC service invocations under the unique temperature, energy, and resource constraints of an OEC satellite, e.g., allowing temporal shifting of computation to times when a satellite can harvest solar energy~\cite{sahraei2023xfaas,schirmer2023profaastinate}.
    \item This level of abstraction also reduces development costs for clients, who no longer have to manage resource scheduling or other operational concerns in their OEC services~\cite{hendrickson2016serverless,jonas2019cloud}.
    \item Shared application runtimes reduce the size of service deployment packages, which helps reduce deployment costs given the high uplink costs for satellites. Instead of uplinking, e.g., a Python runtime per service (as would be required for containers or virtual machines), deploying a FaaS function requires only its code.
    \item The FaaS execution model, where client code is executed in response to events, aligns with the continuous generation of Earth observation events by satellite sensing equipment.
\end{itemize}

We elaborate the opportunities and challenges of shared satellite platforms in \cref{sec:challenges}.
We give an overview of the architecture of \trabant{} in \cref{sec:approach} and evaluate it with a proof-of-concept prototype in \cref{sec:evaluation}.
Specifically, we use real OEC satellite traces and telemetry data to replicate a realistic OEC environment for this evaluation.
Finally, we provide an outlook and avenues for future work in \cref{sec:discussion}.
We make all artifacts developed for this paper available as open-source software.\footnote{\url{https://github.com/project-spencer/trabant}}

\section{Background}
\label{sec:background}

We first give an overview of Earth observation satellites, orbital edge computing, and serverless edge computing.

\subsection{Earth Observation Satellites}

\begin{figure}
    \centering
    \includegraphics[width=\linewidth]{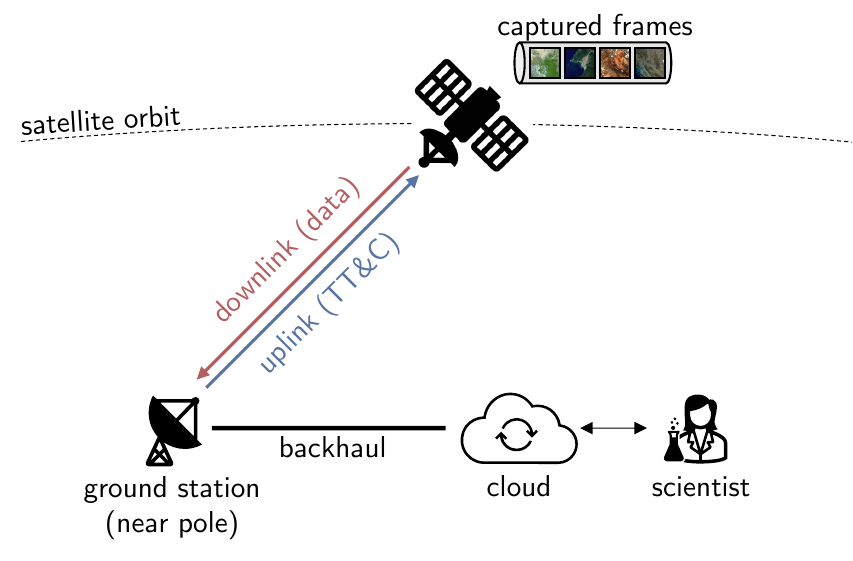}
    \caption{Traditional Earth observation satellite capture images of Earth from LEO and downlink them when passing ground stations, which are usually located near the Earth's poles. Data is sent through the backhaul network to a cloud for processing, from which scientists can access it for analysis}
    \label{fig:satellites}
\end{figure}

Earth observation satellites collect data on Earth's land, oceans, or atmosphere, providing a crucial data source for remote sensing applications in climate science, agriculture, finance, or disaster monitoring.
\qty{95}{\percent} of the more than \num{1200} Earth observation satellites orbiting Earth today are located at altitudes below \qty{2000}{\km}, in low-Earth orbit (LEO)~\cite{ucssd,nasaglossary}.
This low orbit allows high-resolution data capture, limits the power required to downlink data, and is cheaper to launch into~\cite{denby2023kodan}.
To maintain orbit, LEO satellites perform a revolution around Earth every \qtyrange{1}{2}{\hour}.
Many of the Earth observation satellites active today are small, e.g., following the CubeSat standard with a standardized form factor of a \qty{10}{\centi\meter} cube (\qty{1}{\text{U}})~\cite{mehrparvar2022cubesat}, and use commercial-off-the-shelf (COTS) equipment, keeping development costs as low as \$\num{65000}~\cite{denby2020orbital,exakratos,cappaert2018building,bouwmeester2010survey}.

We show a typical Earth observation satellite architecture in \cref{fig:satellites}.
The satellite continuously captures \emph{frames}, i.e., images of Earth in different bands (with different central wavelengths)~\cite{denby2023kodan}.
Satellites periodically pass ground stations that they can downlink data to~\cite{vasisht2021l2d2}.
Such ground stations are usually located near the Earth's poles, as this allows the satellite to perform downlinking once per orbit despite Earth's rotation.
Depending on the satellite's power constraints and antenna size, total downlink data size for a typical Earth observation satellite can range from hundreds of megabytes to a few gigabytes per pass~\cite{vasisht2021l2d2,devaraj2019planet}. 
In the uplink direction, satellites can receive tracking, telemetry, and control (TT\&C) data, e.g., for acknowledgements, although this link is usually narrowband and thus constrained to tens to hundreds of kilobits per second~\cite{vasisht2021l2d2,xing2024deciphering}.

A simple calculation shows that downlink rate is a key bottleneck for Earth observation:
Consider a satellite capturing images at a \qtyproduct{256 x 256}{\pixel} frame resolution with each pixel representing \qtyproduct{10 x 10}{\m} on the ground (a relatively subpar resolution).
Within a single orbit, the satellite would collect $\frac{\qty{40000}{\km}}{\qty{256}{\pixel} \times \qty{10}{\m\per\pixel}} \simeq \num{15628}$ individual frames.
Using only red, green, and blue bands with one byte per pixel, this would yield $\sim \qty{3.07}{\giga\byte}$ of raw data to be downlinked per orbit.

\subsection{Orbital Edge Computing}

Orbital edge computing (OEC) is a new approach to limit the cost of data downlinking through on-board data processing~\cite{denby2020orbital,denby2023kodan,wang2023satellite,xing2024deciphering,leyva2023satellite,wang2023vision,furutanpey2024fool,yin2024comprehensive,furano2020towards,horine2021creating,greene2023system,lei2024do}.
By executing filtering and inference on the satellite, unneeded data can be discarded, reducing the strain on the downlink.
For example, ESA's $\Phi$-SAT-1~\cite{giuffrida2021varphi,giuffrida2020cloudscout} mission has demonstrated on-board cloud segmentation using a machine learning (ML) model.
Beyond that, information can be extracted directly from the raw data to then downlink only an ML inference result instead of a whole frame.
Such filtering and inference not only saves bandwidth, but reduces the time to generate insights from Earth observation data.

Unlike satellite flight controllers, which are critical components, OEC can use commodity COTS computing hardware, such as Raspberry Pi or NVIDIA Jetson boards~\cite{xing2024deciphering,arechiga2018onboard,slater2020total,orbitsedge,spacecloud}.
The limited radiation in LEO may lead to infrequent hardware crashes as a result of single-event upsets (SEU) but has been shown to not affect performance or lifetime for typical missions~\cite{xing2024deciphering,pfandzelter2023failure}.
The main constraints for such hardware are energy availability and heat dissipation~\cite{xing2024deciphering}.
Most Earth observation satellites use a combination of solar panels and batteries~\cite{garcia2021electric,raffaele2023cubesat}.
Depending on solar panel area and sunlight, a CubeSat can harvest on the order of \qtyrange{1}{150}{\watt} of energy at maximum~\cite{garcia2021electric,xing2024deciphering,pumpkin}.
If the satellite is eclipsed by Earth or not ideally positioned for full solar exposure, less or no energy can be harvested.
During this time, the satellite may use its batteries to drive communication and computation equipment.

Given the lack of atmosphere in LEO, satellite on-board computing devices must make use of radiative cooling, which is relatively inefficient especially for smaller satellites with little surface area~\cite{yendler2021thermal,xing2024deciphering}.
Such passive heat dissipation using the satellite structure must also ensure that the operational temperature limits of the satellite are not exceeded, as this could permanently damage satellite components.
Consequently, there is a limit on how much heat a computing chip may generate during its operation.

\subsection{Serverless Edge Computing}

Serverless computing is a popular cloud deployment paradigm that decouples application logic from operational concerns such as elastic scaling~\cite{jonas2019cloud,hendrickson2016serverless}.
Function-as-a-Service (FaaS) is one of the predominant serverless programming models.
In FaaS, applications are composed of small, stateless functions that can be invoked in response to events, such as HTTP requests, asynchronous messages, or changes to a database.
Developers only write the function code using a high-level programming language, while the underlying platform, which is managed by the cloud provider, handles resource allocation, elastic on-demand function instantiation, and scale-to-zero.
This benefits not only application developers, who can outsource operational concerns and focus on business logic, but also cloud platform operators, who can allocate their available resources efficiently and dynamically pool resources between tenants~\cite{pemberton2019serverless}.

At the edge, where resources are more constrained than in a cloud data center, FaaS can have similar benefits~\cite{aslanpour2021serverless,raith2023serverless,denby2020orbital,shi2016edge,xie2021serverless}.
Rather than allocating fixed `slices' of infrastructure for a tenant, an edge service provider can dynamically allocate the limited edge resources between functions of different tenants, creating the illusion that each tenant can always use as many resources as they require.

\section{Multi-Tenant Orbital Edge Computing}
\label{sec:challenges}

Before introducing the \trabant{} architecture, we elaborate how we believe existing satellite mission design falls short of the benefits that OEC has, and how shared OEC satellites could address these limitations by decoupling the satellite platform from the applications and missions running on it.

\paragraph*{Mission Cost}
An obvious disadvantage of planning a separate satellite mission for each application is cost.
Even with COTS equipment, parts costs for a 1U CubeSat can be on the order of \$\num{65000}~\cite{denby2023kodan,exakratos}.
Launch costs for commercial ride-share launches can be on the order of \$\qty[per-mode=symbol]{2500}{\per\kilo\gram}~\cite{villasboas2023innovative}, but additional costs for shipping, installation, payload separation, and operation can increase this to \$\num{100000} per mission~\cite{nieto2019cubesat}.
In addition, planning a satellite mission, building the satellite, and operating it over a multi-year lifetime require significant human resources.
These costs are prohibitive for most Earth observation use-cases, where scientists must instead rely on public commercial Earth imagery providers with lower-resolution data that can be days or weeks old.
\emph{A shared satellite platform could distribute these missions costs between the tenants over the lifetime of the satellite.}

\paragraph*{Mission Lead Times}
Along with mission development costs, satellite missions also have long lead times, including manufacturing the hardware itself, licensing the satellite and communication frequencies, and waiting for a suitable launch opportunity.
CubeSat missions regularly need 1--2 years from start to the satellite being inserted into orbit~\cite{nieto2019cubesat}.
Application requirements regarding sensor resolution, downlink bandwidth, and compute resources have to be known early in this process.
These mission lead times mean that any OEC software meant to provide valuable insights from orbit is already years old by the time it becomes operational, a significant time span compared to advances in software development.
\emph{Decoupling OEC software development from satellite development can reduce lead times for new applications to months or even weeks.}

\paragraph*{Mission/Hardware/Software Alignment}
When planning a satellite mission for a specific application, mission design has to be carefully balanced to manage costs.
As weight is the main factor for launch costs, antenna size, solar panel area, battery capacity, OEC compute equipment, and sensors must be chosen to fit perfectly.
For example, harvesting too little solar power impacts the functionality of the satellite, while harvesting too much means that some launch weight was wasted.
Similarly, the compute equipment must be able to run the filtering or inference logic in the time it takes to capture a new frame, i.e., not lead to queuing, yet executing the logic too quickly implies that OEC equipment was overprovisioned.
Then, the downlink bandwidth (and antenna size and power budget) must fit the expected data size, i.e., the frame resolution, frame capture frequency, and expected data minimization through the OEC software.
While these constraints can of course be solved with careful mission planning, they significantly constrain the flexibility of the satellite.
Updating the mission, e.g., with new filtering logic that is more accurate but requires more compute resources or leads to more total downlinked data, is not possible.
As a result, the data that an Earth observation satellite generates over its lifetime can become less valuable as the mission progresses.
\emph{Running multiple applications on a shared satellite platform with OEC allows flexible allocation and reallocation of the mission to different applications.}

\paragraph*{Development Cost for Constrained Software}
OEC is a unique environment to deploy and manage application services in.
OEC software must deal with varying energy availability, using excess solar power when the satellite is in sunlight and not entirely depleting the battery when the satellite is eclipsed.
At the same time, the software must refrain from sustained resource utilization, as this could increase equipment temperature beyond operational limits.
Finally, while unlikely in LEO, SEU caused by radiation outside the Earth's atmosphere may lock up the OEC computer, necessitating a hard reboot across which the software must maintain functionality.
While not impossible to design for, these constraints are hardly intuitive for those providing the OEC application logic, who are experts in Earth observation or related fields rather than OEC software developers.
Additionally, the effort to develop software to fit these constraints is repeated for every satellite mission, as there are no existing OEC operating systems to build on top of.
\emph{A shared satellite OEC environment could provide a software platform tailored to the satellite, abstracting these constraints for the application services running on it.}

\paragraph*{LEO Satellites Spend Little Time Over Their Regions-of-Interest}
To maintain their orbit, satellites in LEO must move at speeds in excess of \qty[per-mode=symbol]{25000}{\km\per\hour}, while Earth continuously rotates underneath.
Over time, the satellite covers different geographical areas that may not be of interest to its mission.
For example, a satellite monitoring wildfires in Australia from \qty{520}{\km} altitude in a typical sun-synchronous orbit (\ang{97} inclination) will spend only \qty{2.5}{\percent} of its lifetime over Australia~\cite{lu2024onboard,kanyinidb}.
Of course this satellite could be designed to wake up and capture frames only over a specific geographic area, with power, downlink, and compute resources budgeted accordingly, although the sensors could also be used for other missions.
From a cost perspective, fixed development and operational costs remain.
\emph{Serving multiple tenants and applications means that a shared satellite can provide more value as it visits different geographic areas during each orbit.}

\paragraph*{Environmental Impact}
Finally, the environmental impacts of having hundreds or thousands of application-specific satellites in LEO are worth considering.
Each satellite launch releases considerable amounts of greenhouse gases into the atmosphere and contributes to stratospheric ozone depletion~\cite{ryan2022impact}.
During its mission, the satellite requires a portion of LEO, becoming a collision risk for other satellites~\cite{kessler1978collision,boley2021satellite} and possibly contributing to light pollution~\cite{mcdowell2020low}.
Most LEO satellites deorbit after some time as a result of atmospheric drag, yet burning up in the atmosphere is suspected to negatively impact Earth's ozone layer~\cite{ferreira2024potential}.
\emph{Although a shared satellite may be larger than an application-specific one, reducing the overall number and mass of satellites in Earth orbit can reduce the environmental impact.}

\section{\trabant{} Approach}
\label{sec:approach}

\trabant{} is a serverless approach to building a shared OEC satellite software platform.
Clients, such as scientists, provide filtering, inference, and selection logic for the satellite as FaaS functions.
The satellite continuously captures multi-spectral frames of Earth at its given sample distance and rate.
\trabant{} takes these frames and executes the clients' logic against them, providing isolation for client code and ensuring that the constraints of the satellite regarding power consumption and temperature are adhered to.
At the same time, it decouples the effort of managing the complex OEC execution environment from function implementation.

\begin{figure}
    \centering
    \includegraphics[width=\linewidth]{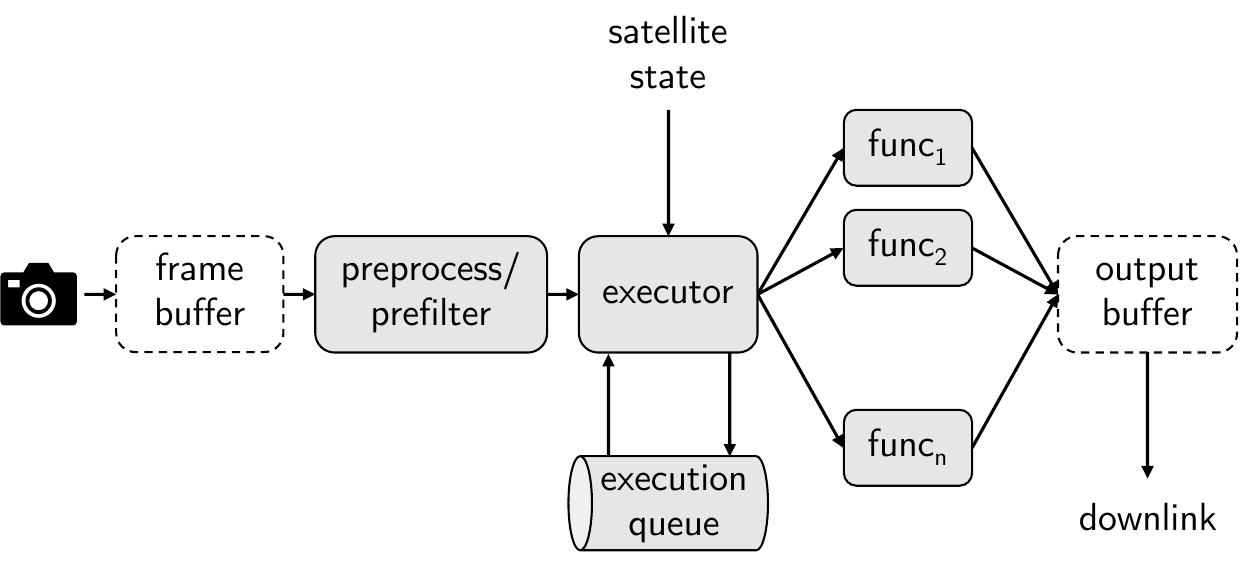}
    \caption{\trabant{} is a serverless architecture for on-board processing of Earth observation frames. Captured frames are preprocessed and prefiltered, with an executor either directly invoking tenants' functions or enqueuing them when temperature or power constraints do not allow processing. The function output, which can be filtered or inferred data, is then buffered for downlinking.}
    \label{fig:trabant}
\end{figure}

We provide an overview of the \trabant{} components in \cref{fig:trabant}.
Captured frames first reach the \emph{frame buffer}, the interface between the camera/sensor subsystem on the satellite and the OEC computer.
Raw frame data must be adjusted for optical distortion before it can be used by an application and atmospheric turbulence or cloud cover could obscure the view of Earth.
As applications expect accurate data without significant cloud cover or distortions~\cite{soret2024semantic,giuffrida2021varphi}, \trabant{} includes a \emph{preprocessing/prefiltering} step.
This step, which runs on the OEC hardware, already filters out frames that do not meet a certain quality level.
Note that this means that the rate at which applications receive frames for processing may be lower than the actual satellite frame capturing rate.

An \emph{executor} invokes the different applications with the frame and its metadata, such as current time or location.
The executor does not necessarily invoke each function instantly:
Instead, it continuously monitors the state of the satellite, including battery charge levels and temperature.
If the available energy is too low or temperature too high, the executor puts invocations (pairs of function and frame references) into a persistent execution queue.
Once sufficient energy is available and the thermal state permits, the queue can be processed.
Time-shifting FaaS invocations in this manner is a well-known technique employed, e.g., in Meta's \emph{XFaaS}~\cite{sahraei2023xfaas,schirmer2023profaastinate}.

Each tenant application runs as a FaaS function in its own isolated process.
An implementation of \trabant{} may provide different language and library runtimes, e.g., for Python or a specific ML library.
The results of each invocation are transferred into an output buffer that can be downlinked once the satellite is in contact with a ground station.
These results could be whole frames that an application deems worthy of downlinking, cropped frames with higher value for the client, or just inference results.

While specific pricing strategies are out of scope for this paper, we envision using the same usage-based pricing strategies used in cloud computing today.
This incentivizes users to limit their resource usage and allows operators to amortize cost.
While the specific pricing strategies are outside the scope of this paper, we discuss avenues for future works in this field in \cref{sec:discussion}.

Beyond ensuring that OEC equipment only performs processing tasks when energy is available and the satellite temperature is not close to its operational limits, note that \trabant{} keeps no state other than the persistent execution queue and output buffer.
As a result, there is no significant impact on applications during an unexpected OEC reset, e.g., caused by SEU or when too much energy is required for communication or satellite attitude control.
Each independent step, including the application functions, is free of side effects and can thus be repeated if necessary.
Note that this still assumes that the actual hard drive, e.g., flash storage, is sufficiently resilient to SEU, which is an orthogonal challenge that is also present without \trabant{}~\cite{chen2017heavy,henkel2023mitigating}.

Of course, simply offering the satellite OEC resources as a platform for tenants does not mean that there are no energy, thermal, computational, or downlink constraints for the entire OEC software system.
These constraints still limit the overall computational capabilities of the satellite and allocating more client services than the satellite can handle will inevitably lead to backpressure.
We believe that the FaaS model significantly simplifies reasoning about the resource consumption of the services running on the shared satellite.
Specifically, we argue for modelling the energy consumption of each function independently by executing it with a sufficiently large (realistic) set of input frames on compute hardware identical to that of the OEC satellite and measuring the average energy required for each invocation.
The high level of abstraction that the function is programmed for makes this trivial.
Further, this should also give insight to the compression ratio of the service, i.e., the number of bytes it outputs for downlinking compared to the number of input bytes (frame count and size).
These two variables must be considered for each function addition or replacement.
Along with some constants specific to the satellite (which can be quantified during satellite development), the following constraint must hold when considering a change to the functions ${f_1, f_2, \dots, f_n}$ on the OEC satellite:

\begin{equation}
    P_{\text{generated}} \geq P_{\text{compute}} + P_{\text{comm}}
\end{equation}

, i.e., the power harvested for compute and communication must be greater than that spent, where:

\begin{equation}
    P_{\text{compute}} = P_{\text{base}} + E_{\text{pre}} \times R_{\text{frame}} + \sum_{i = 1}^{n}{E_{f_i} \times R_{\text{frame}} \times (1 - R_{\text{filter}})}
\end{equation}

Here, $E_{\text{base}}$ is the base power draw of the OEC equipment and \trabant{}, $E_{\text{pre}}$ is the energy required to preprocess a frame, $R_{\text{frame}}$ is the frame capture rate, $R_{\text{filter}}$ is the rate at which frames are discarded by the preprocessing step, and $E_{f_i}$ is the energy required to execute one invocation of $f_i$.
Note that this assumes power consumption to grow linearly with compute utilization.
This is a simplifying assumption but serves as a starting point and are sufficient to investigate the \trabant{} approach.

Communication power budget can be calculated as:

\begin{equation}
    P_{\text{comm}} = E_{\text{sendbyte}} \times \sum_{i = 1}^{n}{C_{f_i}} \times R_{\text{frame}} \times (1 - R_{\text{filter}}) \times S_{\text{frame}}
\end{equation}

Here, $E_{\text{sendbyte}}$ is the energy required to downlink a single byte, $C_{f_i}$ is the compression ratio of $f_i$, and $S_{\text{frame}}$ is the size of a frame in bytes.

Next to energy budget, we must also ensure that the downlink budget is not exceeded:

\begin{equation}
    B_{\text{downlink}} \geq \sum_{i = 1}^{n}{C_{f_i}} \times R_{\text{frame}} \times (1 - R_{\text{filter}}) \times S_{\text{frame}}
\end{equation}

, where $B_{\text{downlink}}$ is the budget for data that can be queued for downlinking, which can be calculated by dividing data rate during ground station contact by contact rate.

Finally, we must ensure that frames can be processed as quickly as they are captured.
Specifically, it must hold:

\begin{equation}
    {R_{\text{frame}}}^{-1} \geq T_{\text{pre}} + \sum_{i = 1}^{n}{T_{f_i}} \times (1 - R_{\text{filter}})
\end{equation}

, where $T_{\text{pre}}$ is the time taken to preprocess a frame and $T_{f_i}$ is the time taken to process one frame with $f_i$.
Note that this ignores possible benefits from parallelization of processing on multicore architectures and thus serves as an upper bound.

\section{Evaluation}
\label{sec:evaluation}

To evaluate \trabant{} empirically, we replicate an OEC satellite in a local testbed, deploy a prototype implementation of \trabant{} on this testbed, and use satellite traces and ML workloads on this system.

\subsection{Evaluation Environment}

\subsubsection{Scenario}

\begin{figure}
    \centering
    \includegraphics[height=1.8in]{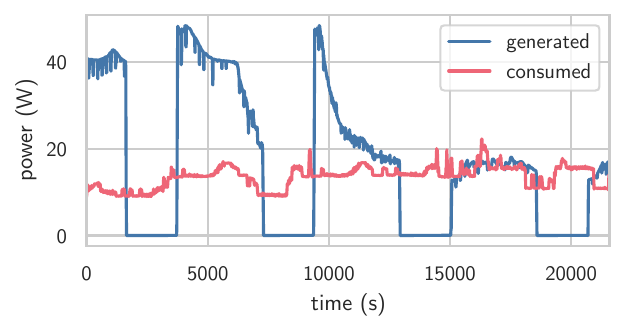}
    \caption{During our 6-hour trace, BUPT-1 consumes a mean \qty{13.51}{\watt} of power, while its solar array generates between \qty{0.00}{\watt} (darkness) and \qty{48.59}{\watt} (maximum at sunlight) of power. Power output is also unstable during sunlit periods, with more power generated at the beginning than at the end, a result of the satellite's angle relative to the sun.}
    \label{fig:energy}
\end{figure}

\begin{figure}
    \centering
    \includegraphics[height=1.3in]{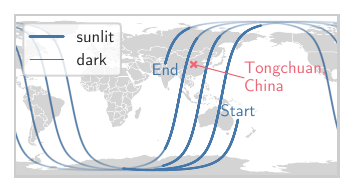}
    \caption{Our 6-hour trace from May 1\textsuperscript{st}, 2023, encompasses four orbits, with the satellite coming into contact with the ground station in Tongchuan toward the end of its third orbit}
    \label{fig:trajectory}
\end{figure}

We base our evaluation on architecture and traces of the BUPT-1 satellite~\cite{xing2024deciphering}.
BUPT-1 is a 12U CubeSat research satellite hosting a Raspberry Pi 4B compute board for general-purpose computation.
Traces for BUPT-1 include generated solar power over time, energy spent on satellite operations, communication and computation, and temperature measurements, all at one-second granularity.
BUPT-1 orbits Earth at altitudes between \qty{487}{\kilo\meter} and \qty{494}{\kilo\meter} with an inclination of \ang{97.3}.
BUPT-1 can harvest up to \qty{40}{\watt} of solar power from two solar arrays and has rechargeable lithium batteries to store energy.
The batteries have a total capacity of \qty{115}{\watt\hour} but must not be discharged further than \qty{30}{\percent}, which would shorten their lifetime.
As the batteries are overprovisioned and designed to power four computers instead of just a single Raspberry Pi, we use a more constrained \qty{57.5}{\watt\hour} battery capacity in our experiments (one fourth of the original capacity).
To communicate with the satellite, a ground station in Tongchuan, China is available with a \qty{1}{\mega\bps} uplink and \qty{100}{\mega\bps} downlink.
We assume a \qty{20}{\percent} protocol and error-correction overhead for each of these links.
We use energy and trajectory traces of BUPT-1 for a 6-hour period on May 1\textsuperscript{st}, 2023.
We show the generated and consumed (by the satellite) power of our 6-hour section of the trace in \cref{fig:energy} and the satellite trajectory in \cref{fig:trajectory}.

We base our evaluation on BUPT-1 as it is the only satellite equipped with COTS computing equipment that has detailed traces openly available with hardware that can be realistically replicated in a lab environment.
Of course, there are more powerful Earth observation satellites in terms of energy harvesting efficiency, compute resources, downlink bandwidth, or ground station pass frequency.
Nevertheless, as we discuss in \cref{sec:discussion}, increasing any of these parameters does not obviate the need to adhere to any constraints when scheduling compute services on an Earth observation satellite, it simply changes them.

\subsubsection{Testbed}

\begin{figure}
    \centering
    \includegraphics[width=\linewidth]{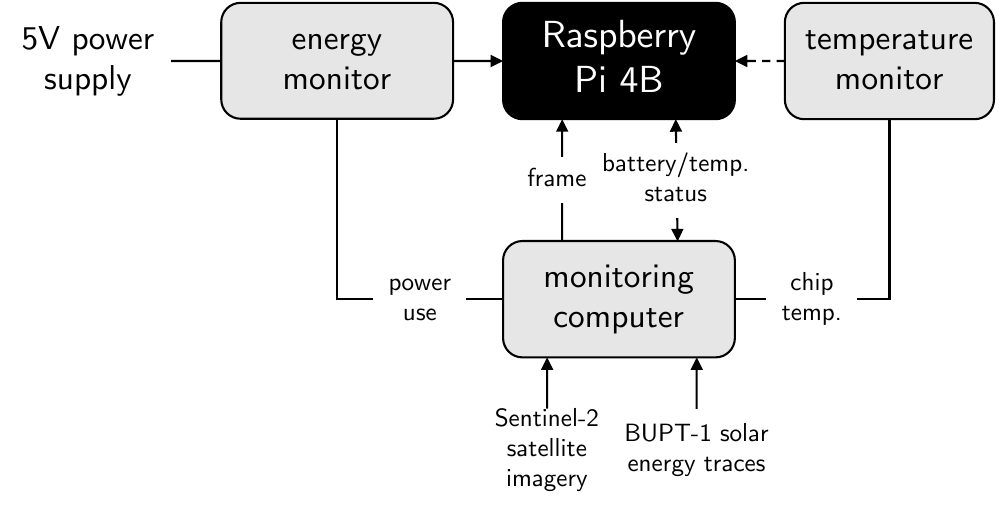}
    \caption{The local testbed architecture we use to evaluate \trabant{} contains a Raspberry Pi 4B hosting our \trabant{} prototype and workloads. A monitoring computer reads energy and temperature of the Raspberry Pi, models battery levels, and sends input frames.}
    \label{fig:evalsetup}
\end{figure}

For our evaluation, we replicate the BUPT-1 environment in a local testbed.
As shown in \cref{fig:evalsetup}, we use a Raspberry Pi 4B as our compute module.
A monitoring computer connected to this Raspberry Pi over Ethernet emulates the satellite environment by monitoring its chip temperature and overall energy consumption.
Using traces from BUPT-1 solar energy harvesting with one-second granularity and measured energy consumption, we model the battery charge levels over time.
This information, along with the current external chip temperature, is provided to our \trabant{} prototype over an HTTP API (in lieu of a real satellite interface that would provide this information).
Note that in their evaluation of the BUPT-1 satellite, Xing et al.~\cite{xing2024deciphering} have shown that a ground-based replica of the on-board Raspberry Pi 4B has similar energy and temperature performance, leading us to believe that our ground-based experiments are representative of performance in LEO.

\subsubsection{Prototype}

Our proof-of-concept prototype of \trabant{} and our monitoring are implemented in Go, loosely based on the \emph{tinyFaaS} edge serverless platform~\cite{pfandzelter2020tinyfaas}.
Frames are sent to our prototype as binary data over HTTP, where frames that are not sunlit are filtered out.
Next, our frames are preprocessed with an implementation of the Braaten-Cohen-Yang pixel-level cloud detection algorithm~\cite{braaten2015automated}, providing an estimate of frame cloud cover.
Frames with a cloud cover larger than \qty{30}{\percent} are not processed further, others are enqueued for processing by our functions.
Our executor dequeues function calls if it is idle, meaning the battery is not discharged more than \qty{30}{\percent} and the external chip temperature is below \qty{50}{\degreeCelsius}.
We use this low temperature limit to approximate the difficulty of heat dissipation in space.
Each function handler is a Docker container using a Python3 runtime and exposing an HTTP endpoint for invocations.
Our Python3 runtime has common ML dependencies such as the TensorFlow Lite runtime pre-installed.
For every frame, each function receives frame metadata, such as latitude, longitude, and cloud cover, and a path to the frame data on a shared logical volume.
The function handler then ensures that any returned data is persisted as output data for downlinking.
We emulate downlinking by reading from the output buffer at the downlink rate when the satellite is in contact with the ground station, modelling antenna power consumption accordingly.

\subsubsection{Workload}

\begin{figure}
    \centering
    \begin{subfigure}[t]{0.249\linewidth}
        \centering
        \includegraphics[width=0.95\linewidth]{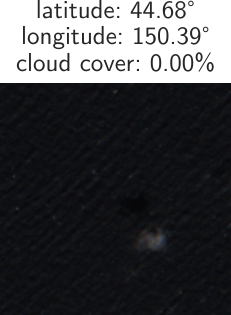}
        \caption{Ocean}
        \label{fig:example_frames:ocean}
    \end{subfigure}%
    \hfill
    \begin{subfigure}[t]{0.249\linewidth}
        \centering
        \includegraphics[width=0.95\linewidth]{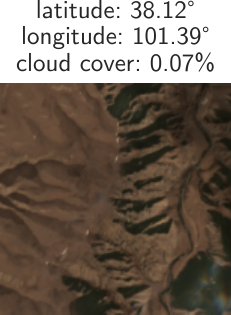}
        \caption{Land}
        \label{fig:example_frames:land}
    \end{subfigure}%
    \hfill
    \begin{subfigure}[t]{0.249\linewidth}
        \centering
        \includegraphics[width=0.95\linewidth]{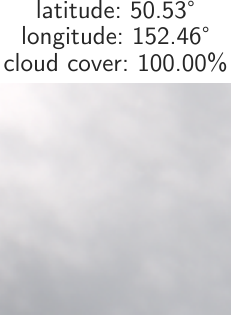}
        \caption{Cloud}
        \label{fig:example_frames:cloud}
    \end{subfigure}%
    \hfill
    \begin{subfigure}[t]{0.249\linewidth}
        \centering
        \includegraphics[width=0.95\linewidth]{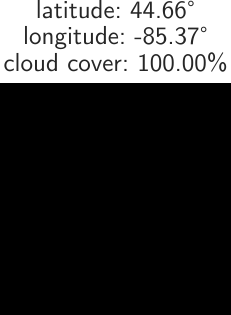}
        \caption{Darkness}
        \label{fig:example_frames:night}
    \end{subfigure}%
    \caption{Example frames of our workload, each covering \qtyproduct{2560 x 2560}{\meter} at \qty[per-mode=symbol]{10}{\meter\per\pixel}. Sunlit frames (\crefrange{fig:example_frames:ocean}{fig:example_frames:cloud}) are from Sentinel-2 satellites~\cite{drusch2012sentinel,sentinel2}, dark frames are from VIIRS imagery~\cite{viirs,viirsreport}.}
    \label{fig:example_frames}
\end{figure}

Our monitoring computer provides a constant stream of ``captured frames'' as an input workload.
This trace is based on \emph{Sentinel-2}~\cite{drusch2012sentinel,sentinel2} and \emph{Visible Infrared Imaging Radiometer Suite} (VIIRS)~\cite{viirs,viirsreport} satellite imagery with the 13 Sentinel-2 bands.
At a ground resolution of \qty[per-mode=symbol]{10}{\meter\per\pixel} and frame size of \qtyproduct{256 x 256}{\pixel}, the satellite would capture a new frame every \qty{400}{\milli\second}.
For each of those frame areas, we first calculate if the image would be in darkness or sunlight.
For dark images (\qty{59.2}{\percent}), we use night data captured by the VIIRS instrument.
For sunlit frames (\qty{40.8}{\percent}), we download the most recent image of the frame location before May 1\textsuperscript{st}, 2023, or use a random ocean frame (Sentinel-2 does not provide acquisitions of most of the world's ocean area).
We show example frames in \cref{fig:example_frames}.
Preliminary analysis of our collected sunlit frames with the \emph{s2cloudless}~\cite{s2cloudless} ML cloud detector shows a total cloud fraction of \qty{36.3}{\percent}, which is smaller than expected in reality (mean $\sim\qty{67}{\percent}$ world cloud cover~\cite{king2013spatial}), but ensures that our workload size is not favored towards our experiments.
In total, \qty{44.4}{\percent} of our frames have a cloud cover larger than \qty{30}{\percent}.

\subsubsection{Functions}

\begin{table}
    \caption{Input Parameters for Our Scenario and Prototype}
    \label{tab:measured}
    \centering
    \begin{tabular}{
            l
            r
            l
            l
        }
        \toprule
        \textbf{Variable}      & \multicolumn{2}{c}{\textbf{Value}} & \textbf{Source}                                                   \\
        \midrule
        $P_{\text{generated}}$ & \num{2950}                         & \unit{\milli\watt}                              & Satellite Trace \\
        $P_{\text{base}}$      & \num{1518}                         & \unit{\milli\watt}                              & Measured        \\
        $E_{\text{sendbyte}}$  & \num{2}                            & \unit[per-mode = symbol]{\micro\joule\per\byte} & Satellite Info  \\
        $E_{\text{pre}}$       & \num{0.01}                         & \unit{\joule}                                   & Measured        \\
        $R_{\text{frame}}$     & \num{2.5}                          & \unit{\hertz}                                   & Satellite Trace \\
        $R_{\text{filter}}$    & \num{77.0}                         & \unit{\percent}                                 & Frame Trace     \\
        $S_{\text{frame}}$     & \numproduct{256 x 256 x 13}        & \unit{\byte}                                    & Frame Trace     \\
        $B_{\text{downlink}}$  & \num{600}                          & \unit{\kilo\bps}                                & Satellite Trace \\
        $T_{\text{pre}}$       & \num{0.038}                        & \unit{\second}                                  & Measured        \\
        \bottomrule
    \end{tabular}
\end{table}

We use five ML functions as our on-board workloads, each implemented using Python and TensorFlow.
Each model performs ML inference on single satellite frames using the TensorFlow Lite runtime.
Note that while all ML models we use are trained on real data and able to perform their respective tasks with acceptable accuracy, the goal of our functions is to present a realistic workload rather than the best possible model performance.
As such, our models are also designed to have small footprints (model weights on the order of \qtyrange{0.5}{2}{\mega\byte}) in order to run on constrained hardware.

\paragraph*{Methane leak preprocessing}
The goal of the \texttt{methane} function is to classify satellite images to filter out those that do not show industrial areas.
The classification is performed by a convolutional neural network (CNN) trained on the \emph{EuroSat}~\cite{helber2019eurosat} dataset and based on the RGB bands of the image.
If an image has been classified as showing an industrial area, the RGB bands along with two short wave infrared bands (helpful for identifying methane leaks~\cite{varon2021high}) are stored for downlinking.

\begin{table*}
    \caption{Measured Parameters for Our Workload Functions Across Three Repetitions (Experiments Use Values of Repeat 2)}
    \label{tab:functions}
    \centering
    \begin{tabular}{crrrrrrrrr}
        \toprule
                          & \multicolumn{3}{c}{\textbf{$E_{f}$}} & \multicolumn{3}{c}{\textbf{$T_{f}$}} & \multicolumn{3}{c}{\textbf{$C_{f}$}}                                                                                                                                                 \\
        \textbf{Function} & \multicolumn{1}{c}{1}                & \multicolumn{1}{c}{2}                & \multicolumn{1}{c}{3}                & \multicolumn{1}{c}{1} & \multicolumn{1}{c}{2} & \multicolumn{1}{c}{3} & \multicolumn{1}{c}{1} & \multicolumn{1}{c}{2} & \multicolumn{1}{c}{3} \\
        \midrule
        \texttt{methane}  & \qty{0.038}{\joule}                  & \qty{0.041}{\joule}                  & \qty{0.036}{\joule}                  & \qty{0.019}{\second}  & \qty{0.018}{\second}  & \qty{0.018}{\second}  & \qty{0.049}{}         & \qty{0.051}{}         & \qty{0.045}{}         \\
        \texttt{moisture} & \qty{0.107}{\joule}                  & \qty{0.092}{\joule}                  & \qty{0.092}{\joule}                  & \qty{0.050}{\second}  & \qty{0.050}{\second}  & \qty{0.049}{\second}  & \qty{0.029}{}         & \qty{0.041}{}         & \qty{0.018}{}         \\
        \texttt{segment}  & \qty{0.663}{\joule}                  & \qty{0.755}{\joule}                  & \qty{0.826}{\joule}                  & \qty{0.429}{\second}  & \qty{0.494}{\second}  & \qty{0.528}{\second}  & \qty{0.043}{}         & \qty{0.047}{}         & \qty{0.051}{}         \\
        \texttt{vessel}   & \qty{1.075}{\joule}                  & \qty{1.053}{\joule}                  & \qty{1.017}{\joule}                  & \qty{0.562}{\second}  & \qty{0.585}{\second}  & \qty{0.566}{\second}  & \qty{0.024}{}         & \qty{0.026}{}         & \qty{0.021}{}         \\
        \texttt{wildfire} & \qty{0.709}{\joule}                  & \qty{0.667}{\joule}                  & \qty{0.719}{\joule}                  & \qty{0.356}{\second}  & \qty{0.353}{\second}  & \qty{0.356}{\second}  & \qty{0.011}{}         & \qty{0.026}{}         & \qty{0.022}{}         \\
        \texttt{no-op}    & \qty{0.007}{\joule}                  & \qty{0.007}{\joule}                  & \qty{0.010}{\joule}                  & \qty{0.004}{\second}  & \qty{0.004}{\second}  & \qty{0.005}{\second}  & \qty{0.000}{}         & \qty{0.000}{}         & \qty{0.000}{}         \\
        \bottomrule
    \end{tabular}
\end{table*}

\paragraph*{Soil moisture preprocessing}
The \texttt{moisture} function selects images that are predicted to show different agricultural areas, such as vineyards, based on a CNN trained on the \emph{BigEarthNet}~\cite{sumbul2019bigearthnet} dataset.
This CNN uses twelve input bands to perform multilabel classification, i.e., an image can show more than one of the classes.
If the frame is relevant, the RGB bands along with two visible and near infrared bands and one short wave infrared band are stored for downlinking, as these bands can be used to infer soil moisture in subsequent processing~\cite{hegazi2023prediction}.

\paragraph*{Segmentation in China}
The \texttt{segment} function only processes frames that show China based on a simple la\-ti\-tude/\-long\-i\-tude bounding box.
If a frame falls within the bounding box, the function uses a CNN model based on the \emph{U-Net} architecture~\cite{ronneberger2015u} to perform pixel-level segmentation, identifying features such as buildings or roads.
The model is trained on a dataset of segmented satellite images of Dubai~\cite{dubaisegment}.
For every processed frame, the function stores the RGB bands and pixel mask for downlinking.

\paragraph*{Vessel recognition}
The \texttt{vessel} function contains a CNN model trained on the \emph{MASATI}~\cite{gallego2018automatic,alashhab2019precise} dataset and predicts whether a processed frame contains a boat or not.
Before performing inference, the function filters out frames that have a cloud cover larger than \qty{10}{\percent}, which could lead to inaccurate inference results.
If the model decides that the frame may contain a boat with more than \qty{80}{\percent} probability, the RGB bands of the frame are saved for downlinking.

\paragraph*{Wildfire risk detection}
The \texttt{wildfire} function uses a CNN model trained on the \emph{Wildfire Prediction Dataset}~\cite{wildfire} to identify whether the area in a frame is susceptible to wildfires or not.
If the probability for a frame is higher than \qty{60}{\percent}, all bands of the image are stored for downlinking, enabling further analysis on the ground.

\subsection{Analysis}

Before deploying our prototype and workload on our testbed, we explore if and when the constraints proposed in \cref{sec:approach} hold for our workload.
\cref{tab:measured} shows the variables of our particular satellite environment and implementation, calculated from the satellite and image traces or measured in our prototype.
In the full BUPT-1 energy monitoring trace, there is a mean \qty{2950}{\milli\watt} available power, the difference between a mean \qty{15.59}{\watt} solar power and \qty{12.64}{\watt} power draw for satellite upkeep (without payload or data transfer).

Further, from a \qty{80}{\mega\bps} downlink rate and \qty{20}{\watt} antenna power, we arrive at \qty[per-mode=symbol]{2}{\micro\joule\per\byte} energy required to downlink one byte.
From the satellite's trajectory over time, we gather that it is in contact with its ground station for only \qty{0.75}{\percent} of its time in orbit, making the downlink budget $B_{\text{downlink}} = \qty{600}{\kilo\bps}$.
The mean filter rate $R_{\text{filter}}$ is a result of the \qty{40.8}{\percent} sunlit frames of which \qty{44.4}{\percent} have an estimated cloud cover of less than \qty{30}{\percent}.
We measure $P_{\text{base}}$, $E_{\text{pre}}$, and $T_{\text{pre}}$ on our testbed using a no-op function and our energy monitor.

We use a similar approach to quantify function parameters shown in \cref{tab:functions}.
On our testbed, we deploy each function in isolation and send \qty{500}{} randomly selected frames (only sunlit frames with cloud cover lower than \qty{30}{\percent}) at a reduced rate (every \qty{2}{\second}), measuring energy required to process each frame (measured power for the duration of the call minus the base power consumption), time to process each call, and compression ratio of each function (output data size divided by input data size).
For reference, we also include data on a no-op function.
Note that this approach does not require any knowledge about the inner workings of a function:
As we only measure the mean time or energy use across many individual calls, skewed distributions in execution behavior can be accounted for.
For example, despite the \texttt{segment} function ignoring frames outside of China (completion within \qty{4}{\milli\second}) and executing complex inference for the rest (completion time within $\sim$\qty{3}{\second}), we receive a stable mean result across three independent repetitions of the measurement.
We use the results of the second repetition of each measurement for the remainder of this experiment.

Based on the constraints we set in \cref{sec:approach}, we can now calculate which functions we may deploy as our workload.
We find that both our power budget $P_{\text{compute}} = \qty{3.02}{\watt} > P_{\text{generated}}$ and downlink budget $\qty{746.01}{\kilo\bps} > B_{\text{downlink}}$ are exceeded when considering deploying all five functions.
We select the \texttt{methane}, \texttt{moisture}, \texttt{vessel}, and \texttt{wildfire} functions, which have a $P_{\text{compute}} = \qty{2.59}{\watt}$, $P_{\text{comm}} = \qty{0.14}{\watt}$, and total processing time of $\qty{0.266}{\second} < R_{\text{frame}}^{-1} = \qty{0.4}{\second}$, and downlink \qty{564.34}{\kilo\bps}.

\subsection{End-to-End Demonstration}

\begin{figure*}
    \centering
    \begin{subfigure}[t]{0.49\linewidth}
        \centering
        \includegraphics[height=1.8in]{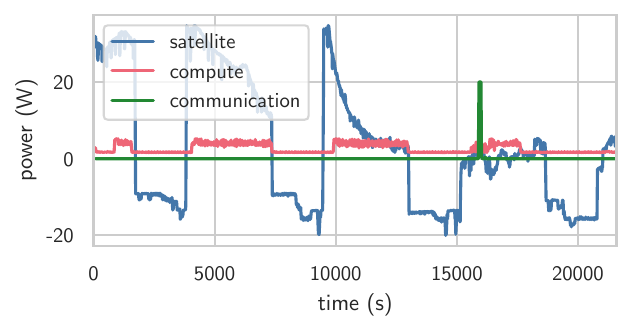}
        \caption{Generated and consumed power}
        \label{fig:full:power}
    \end{subfigure}%
    \hfill
    \begin{subfigure}[t]{0.49\linewidth}
        \centering
        \includegraphics[height=1.8in]{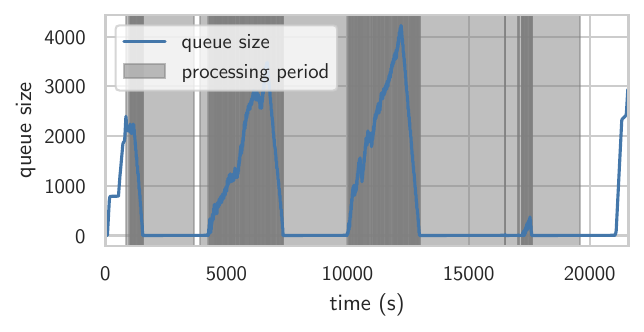}
        \caption{Queue size and processing periods}
        \label{fig:full:queue-size}
    \end{subfigure}%
    \vfill
    \begin{subfigure}[t]{0.49\linewidth}
        \centering
        \includegraphics[height=1.8in]{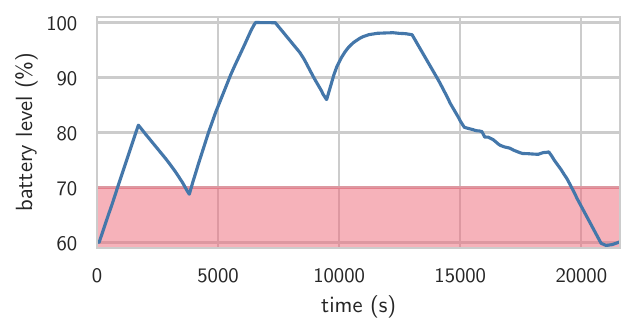}
        \caption{Battery levels}
        \label{fig:full:battery}
    \end{subfigure}%
    \hfill
    \begin{subfigure}[t]{0.49\linewidth}
        \centering
        \includegraphics[height=1.8in]{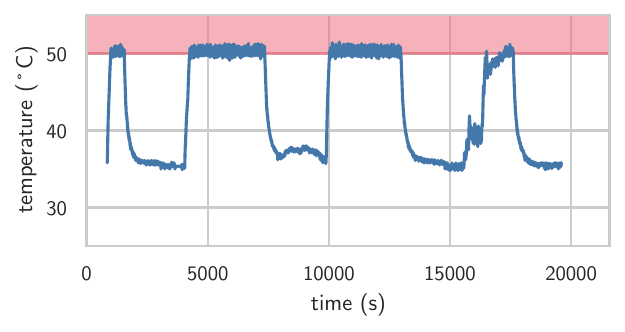}
        \caption{Chip temperature readings}
        \label{fig:full:temp}
    \end{subfigure}%
    \caption{Results of the 6-hour experiment show how \trabant{} time-shifts frame processing under energy and temperature constraints}
    \label{fig:full}
\end{figure*}

We first take a high-level look at the performance of the \trabant{} approach using an end-to-end demonstration of our prototype with the full 6-hour satellite trace.
We use the \texttt{methane}, \texttt{moisture}, \texttt{vessel}, and \texttt{wildfire} functions and start with \qty{60}{\percent} battery capacity and an empty queue.

We show the results of this experiment in \cref{fig:full}.
\cref{fig:full:power} shows the power generation and consumption of our satellite, communication, and computation.
We observe that computing is mostly performed during periods of net positive satellite power, as those are sunlit periods that yield both high solar panel output and sunlit frames for processing.
The satellite comes into contact with its ground station \qty{15189}{\second} into the experiment, when it starts downlinking data for \qty{270}{\second}, with communication power increasing to \qty{20}{\watt} accordingly.
The battery level shown in \cref{fig:full:battery} are mostly affected by satellite power, although we see a small dip when communication starts.
As there is little solar power during the last sunlit period in our trace, the battery charge drops below \qty{70}{\percent} towards the end of our experiment.
We see this reflected in queue size and processing periods in \cref{fig:full:queue-size}, where \trabant{} stops processing new frames (instead queuing them for later), despite a positive power output of the solar panels.
Here, we also see the impact of external chip temperature, as shown in \cref{fig:full:temp}.
Temperature increases as \trabant{} starts processing function calls for incoming frames.
Although there is sufficient energy available for the majority of our experiment, \trabant{} dequeues function calls only until the chip temperature limit of \qty{50}{\degreeCelsius} is reached.
Processing these calls increases chip temperature further to a maximum \qty{51.5}{\degreeCelsius}, yet \trabant{} waits until the compute device has cooled enough to dequeue more invocations.

\subsection{Exceeding Constraints}

\begin{figure*}
    \centering
    \begin{subfigure}[t]{0.32\linewidth}
        \centering
        \includegraphics[height=1.6in]{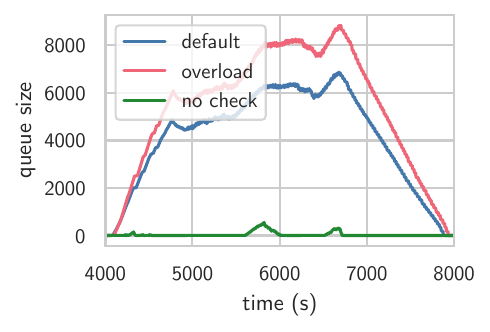}
        \caption{Queue size and processing periods}
        \label{fig:compare:queue-size}
    \end{subfigure}%
    \hfill
    \begin{subfigure}[t]{0.32\linewidth}
        \centering
        \includegraphics[height=1.6in]{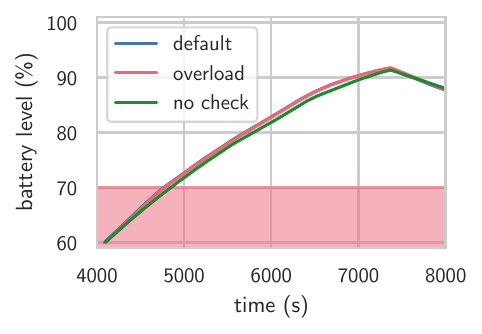}
        \caption{Battery level}
        \label{fig:compare:battery}
    \end{subfigure}%
    \hfill
    \begin{subfigure}[t]{0.32\linewidth}
        \centering
        \includegraphics[height=1.6in]{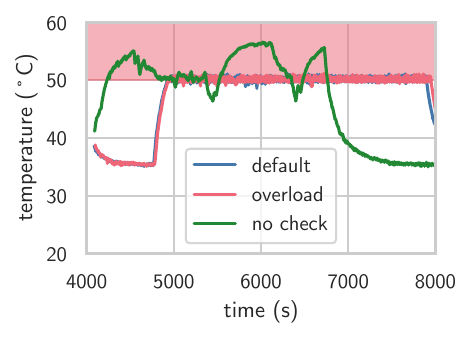}
        \caption{Chip temperature readings}
        \label{fig:compare:temp}
    \end{subfigure}%
    \caption{Results for a \qty{4000}{\second} experiment with additional comparison configurations. `overload' adds the \texttt{segment} function, while `no check' disables the temperature and battery level checks.}
    \label{fig:compare}
\end{figure*}

To further explore how \trabant{}, we perform additional experiments using a subset of our trace starting at \qty{4000}{\second}, shortly before the start of the first full sunlit period.
We compare the same setup (`default') as in the full experiment with two additional configurations:
In the `overload' configuration we add the \texttt{segment} function to our workload.
In the `no-check' configuration, we disable compute time-shifting by removing the temperature and battery level checks.
Further, we start the experiment with only \qty{60}{\percent} battery charge to better observe the impact of different configurations.

Removing the battery level and temperature checks has significant impact on those values, as shown in \cref{fig:compare}.
Although the battery levels are mostly influenced by satellite power, recovering from the low battery charge takes \qty{10}{\percent} longer without state checking, as the system immediately starts processing frames rather than waiting for available power.
This is also reflected in chip temperature, which grows beyond the \qty{50}{\degreeCelsius} limit.
On this small experiment timescale, adding the \texttt{segment} function mainly impacts queue size, with little impact on chip temperature.

\subsection{SEU Resilience}

\begin{figure}
    \centering
    \includegraphics[height=1.6in]{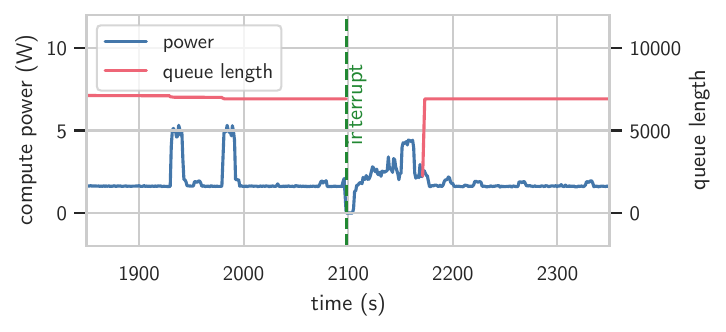}
    \caption{\trabant{} seamlessly resumes processing after a hard reset, e.g., after an interrupt through SEU}
    \label{fig:reset-interrupt}
\end{figure}

Using COTS hardware for satellite computing increases the risk of single-event upsets (SEU) locking up the compute equipment during operation, requiring a hard reset.
In \trabant{}, the FaaS paradigm makes SEU resilience particularly straightforward:
as the functions itself are stateless, we simply persist the input invocation queue to find out which software has processed which frames.
To evaluate SEU resilience of our \trabant{} prototype, we run a subset of our experiment and unplug the Raspberry Pi 4 power cable a few minutes into the experiment, essentially hard resetting the device without signalling the operating system.
We then plug the cable back in to resume operation.

The compute power draw and queue length observations shown in \cref{fig:reset-interrupt} show how \trabant{} dequeues some function invocations shortly before our interrupt (normal operation), with power draw increasing accordingly.
When we cause the interrupt at \qty{2100}{\s}, power draw drops to \qty{0}{\watt}.
As soon as we plug the power cable back in, the device boots and resumes its invocation queue.

\subsection{Deployment Size}

\begin{figure}
    \centering
    \includegraphics[height=1.6in]{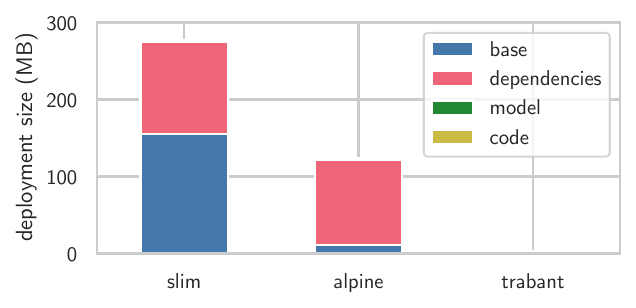}
    \caption{Deployment sizes for the \texttt{moisture} model and code as Docker containers with `slim' and Alpine Linux base images and in \trabant{}}
    \label{fig:upload_sizes}
\end{figure}

In terms of communication, Earth observation satellites are optimized for downlinking data rather than uplinking new commands, which occurs infrequently.
BUPT-1 has a \qty{1}{\mega\bps} uplink, which leads to a theoretical maximum upload size of \qty{75}{\mega\byte} per day, assuming a perfect \qty{10}{\minute} contact window each day.
In reality, this uplink requires error correction and is shared with critical data such as satellite commands.
For a shared satellite platform, deployment sizes for tenant software are thus limited.

\cref{fig:upload_sizes} shows the deployment size of the \texttt{moisture} function in \trabant{} and as a traditional container image.
The model used in the \texttt{moisture} function is the largest of all of our models at \qty{2.77}{\mega\byte}, but this size is dwarfed by the size of the required base images and dependencies.
The official ``slim'' Python3 base image from Docker Hub has a size of \qty{155.47}{\mega\byte}, while dependencies for model inference are an additional \qty{119}{\mega\byte}.
A custom image based on Alpine Linux (which requires manually compiling dependencies such as TensorFlow Lite) has a total size of \qty{121.57}{\mega\byte}.
After compression, the slim and Alpine images are \qty{100.47}{\mega\byte} and \qty{39.42}{\mega\byte} large, respectively.

\trabant{} also requires a base runtime to exist on the satellite, yet runtime files are identical across tenant functions, allowing us to reduce deployment sizes considerably.
Despite this, \trabant{} also allows adding custom dependencies, such as a specific Python library, to deployment packages.
Of course, a similar effect could be achieved by coordinating a shared base image among tenants, which would only need to be uploaded once, yet this would increase complexity for service developers.

\section{Discussion \& Future Work}
\label{sec:discussion}

Our evaluation of the \trabant{} approach has shown that a serverless architecture is well-suited for shared OEC satellites.
We now discuss limitations of our work and derive avenues for future work.

\paragraph*{Earth Observation Satellite Constraints}

Our evaluation of \trabant{} is based on traces of the BUPT-1 satellite.
We chose this evaluation scenario as it can be realistically replicated in a lab based on COTS hardware and available finely-grained, comprehensive traces of satellite trajectory and power consumption.
Of course, BUPT-1 is primarily a research satellite and can thus not compete with state-of-the-art deployed LEO satellites in terms of Earth observation performance.
For example, Planet's High Speed Downlink 2~\cite{devaraj2019planet} achieves a $\num{16.25}\times$ higher downlink rate than we assume for BUPT-1 (\qty{1.3}{\giga\bps} compared to \qty{80}{\mega\bps}) while performing more frequent passes over ground stations further North than Tongchuan, China.
Similarly, Planet's \emph{SkySat} constellation promises a ground resolution on the order of \qty[per-mode=symbol]{0.5}{\m\per\pixel}~\cite{planet2023product}, a $\num{20}\times$ higher resolution than that of the Sentinel-2 data used in our evaluation.
Finally, more efficient solar arrays, e.g., \qty{135}{\watt}~\cite{pumpkin}, are available and more powerful compute resources, e.g., multicore 2nd Generation Intel Xeon Scalable processors with multiple gigabytes of memory on the HPE \emph{Spaceborne Computer-2}~\cite{speed2021hpe,nasa2025cutting}, have been deployed to LEO.
Yet, we note that increasing such parameters does not alleviate the pressure to efficiently downlink data to Earth.
With a tenfold increase in compute performance and downlink bandwidth while resolution and service resource requirements also increases tenfold, the entire satellite system would still require the same efficient scheduling as in our evaluation.
With \trabant{} we make no assumptions on the absolute resource constraints, instead we target the relative mismatch between available resources and the desire to run as much processing as possible on as data of the highest possible resolution.

\paragraph*{Constellations}

Commercial Earth observation satellite operators usually operate tens or hundreds of satellites rather than just one.
Constellations of similar satellites allow more frequent frame captures and higher downlink rates, and we imagine that the \trabant{} approach could also be extended to such constellations.
Specifically, given a set of tenant services and a pool of satellites to host these services, we could use the constraints for each satellite set out in \cref{sec:approach} to calculate service assignments.
A service could also be assigned to multiple satellites in different orbits to yield higher visiting frequency.
While we focus our evaluation on just a single satellite in this work, we plan to explore the challenges of such assignments in future work.

\paragraph*{Trust Model}

Running a service on shared infrastructure raises questions of trust.
Specifically, as is also the case in cloud computing, tenants must trust the operator who runs their services.
Tenants that process sensitive or otherwise critical data may thus not be able to benefit from the shared satellite approach.
We believe that research on trusted serverless computing~\cite{brenner2019trust,shamendra2023trufaas,zhang2022lightweight} can be directly applied here.
Conversely, the operator can largely treat the tenant service as an untrusted workload given the sandboxing in our serverless environment.
Functions cannot access data outside their sandbox and can be monitored if they behave abnormally, e.g., consuming too many CPU cycles, as is standard practice in cloud serverless computing~\cite{agache2020firecracker}.

\paragraph*{Service Isolation}

Similarly to trust between operator and tenant, we must also consider isolation of services between tenants, a pressing research challenge in serverless computing.
Our \trabant{} prototype uses Docker containers for service management and isolation, yet we are aware that this is associated with security risks~\cite{agache2020firecracker,gvisor-gcf}.
Alternative sandboxing mechanisms from research on serverless edge computing, such as microVMs~\cite{agache2020firecracker} or unikernels~\cite{gehberger2022cooling,moebius2024unikernel}, may provide higher levels of isolation at the cost of efficiency and should be explored for our use-case in future work.

\paragraph*{Resource Pricing}

We consider the pricing of functions on a \trabant{} satellite out of scope for this paper, but find it an interesting avenue for future work.
As a starting point, it may be useful to adopt pricing models of cloud serverless compute platforms, which charge for processing time and egress bandwidth~\cite{elgamal2018costless} (with prices adjusted to capital and operational expense of hosting services on a LEO satellite, of course).
We believe that this may also encourage efficient service implementations.

\paragraph*{Base Resource Consumption}

In our experiments, we found that the base power consumption of the Raspberry Pi 4 (\qty{1518}{\milli\watt}) consumed over half of the available \qty{2950}{\milli\watt} generated power, a result of the Raspberry Pi not supporting hibernation.
Given our approach of time-shifted serverless computing in \trabant{}, it might instead be possible to power off the device during periods of low power supply and resume operation later, which we will explore in future research.

\section{Related Work}
\label{sec:relwork}

Orbital edge computing has drawn considerable research interest~\cite{yin2024comprehensive,wang2023vision}.
For example, Giuffrida et al.~\cite{giuffrida2021varphi,giuffrida2020cloudscout} have demonstrated satellite on-board cloud segment using a CNN model running on an Intel Movidius Myriad 2 hardware accelerator on the $\Phi$-Sat-1 satellite.
Further, Denby and Lucia~\cite{denby2020orbital} introduce an OEC architecture for Earth observation missions that parallelizes data collection and processing across a constellation of nanosatellites, which they evaluate with a CNN model identifying building footprint.
With \emph{Kodan}, Denby et al.~\cite{denby2023kodan} present an approach for ML inference in space using a selection of pre-trained models optimized to operate under the constraints of a target satellite, with the runtime dynamically selecting the best model based on geospatial contexts.
Furutanpey et al.~\cite{furutanpey2024fool} present \emph{FOOL}, an OEC-native feature compression method that leverages inter-tile dependencies to reduce the downlink data size of Earth observation frames in a task-agnostic manner.
These approaches are excellent examples of the potential of OEC to reduce the downlink data size in LEO satellites.
However, they also illustrate that the focus of most existing OEC research is to reduce the downlink rate of indiscriminate frame collection, e.g., by filtering out pixels obscured by clouds, or to support a single application, i.e., designing the satellite mission to perform a single kind of inference task.
We believe that shared satellite platforms offer a more flexible way of leveraging OEC but have not been sufficiently studied.

Lei and Saeed~\cite{lei2024do} lay out their vision for constellation-as-a-service satellites with heterogeneous compute resources to support a broad range of customer applications.
Their proposed system uses containers to host services and allocates resources based on geographical position of the satellite.
As no implementation of this system exists at the moment, comparability remains limited.
As we have shown with \trabant{}, a serverless approach requires minimal configuration from customers beyond writing function code and eases deployment logic under the myriad of constraints on an OEC-equipped satellite.

O'Donnel et al.~\cite{o2023extension} extend Amazon Web Services (AWS) edge computing and networking to a LEO satellite in order to seamlessly manage OEC software as if it was running in a traditional cloud.
They use AWS IoT Greengrass for lightweight container orchestration and ML inference with Amazon SageMaker, and evaluate their approach with applications such as fire detection and cloud masking.
This approach allows for flexible reconfiguration of the satellite mission by updating models or applications, yet does not allow sharing of the satellite between different tenants.

In the context of the BUPT-1 satellite and the larger \emph{Tiansuan} constellation of OEC research satellites~\cite{wang2021tiansuan}, Wang et al.~\cite{wang2023satellite} present a study of cloud-native satellites.
The authors argue that using cloud-native technologies, e.g., containers and Kubernetes, significantly simplifies the development and operation of OEC satellites and software.
Their proposed cloud-native satellite architecture allows flexibly allocating a virtualized pool of satellite resources, such as storage, CPU, or hardware accelerators, among different tenant application functions.
This closely aligns with the goals of our work, yet we believe that containers are not well-suited for OEC given deployment sizes and inflexibility in resource allocation.

We have proposed the use of serverless computing for satellite computing in previous work~\cite{pfandzelter2021towards,pfandzelter2024komet} in the slightly different context of LEO in-network computing.
Here, compute resources are embedded in LEO satellite communication constellations (instead of Earth observation satellites) to serve clients on the ground with low-latency access to compute services.
In this context, stateless serverless functions can be seamlessly migrated across interconnected satellites to maintain a static location from the perspective of a ground observer.
While related to this work, communications constellations differ in size, use-case, and capacity to the Earth observation satellites we target with \trabant{}.

\section{Conclusion}
\label{sec:conclusion}

Orbital edge computing reduces the data transmission needs of Earth observation satellites by processing sensor data on-board, allowing near-real-time insights while minimizing downlink costs.
Current orbital edge computing architectures are inflexible and require intricate mission planning that involves upfront knowledge of the OEC software.

In this paper, we have argued for shared satellite platforms that allow multiple clients to execute custom code on-board the satellite to filter and analyze sensor data.
To enable running unknown code from multiple tenants on satellite hardware with varying power and temperature constraints, we leverage the Function-as-a-Service paradigm.
We present \trabant{}, a serverless satellite software platform that reduces operational complexities using automated time-shifted computing.
We demonstrate its capabilities with a proof-of-concept prototype and evaluate it using real satellite computing telemetry data.
Our findings suggest that \trabant{} reduces mission planning overheads, offering a scalable and efficient platform Earth observation missions.
Future work on \trabant{} includes investigating deployment optimization and evaluation on real satellite hardware.

\begin{acks}
    We thank Ruolin Xing, Mengwei Xu, and Shangguang Wang from Beijing University of Posts and Telecommunications for their continued support in understanding the intricacies of the BUPT-1 satellite.

    Partially funded by the \grantsponsor{BMBF}{Bundesministerium für Bildung und Forschung (BMBF, German Federal Ministry of Education and Research)}{https://www.bmbf.de/bmbf/en} in the scope of the Software Campus 3.0 (Technische Universit\"at Berlin) program -- \grantnum{BMBF}{01IS23068}.
\end{acks}

\balance

\bibliographystyle{ACM-Reference-Format}
\bibliography{bibliography.bib}


\begin{thebibliography}{84}


\ifx \showCODEN    \undefined \def \showCODEN     #1{\unskip}     \fi
\ifx \showISBNx    \undefined \def \showISBNx     #1{\unskip}     \fi
\ifx \showISBNxiii \undefined \def \showISBNxiii  #1{\unskip}     \fi
\ifx \showISSN     \undefined \def \showISSN      #1{\unskip}     \fi
\ifx \showLCCN     \undefined \def \showLCCN      #1{\unskip}     \fi
\ifx \shownote     \undefined \def \shownote      #1{#1}          \fi
\ifx \showarticletitle \undefined \def \showarticletitle #1{#1}   \fi
\ifx \showURL      \undefined \def \showURL       {\relax}        \fi
\providecommand\bibfield[2]{#2}
\providecommand\bibinfo[2]{#2}
\providecommand\natexlab[1]{#1}
\providecommand\showeprint[2][]{arXiv:#2}

\bibitem[dub(2020)]%
        {dubaisegment}
Humans in the Loop \bibinfo{year}{2020}\natexlab{}.
\newblock \bibinfo{booktitle}{\emph{Semantic segmentation of aerial imagery}}.
\newblock Humans in the Loop.
\newblock
\urldef\tempurl%
\url{https://www.kaggle.com/datasets/humansintheloop/semantic-segmentation-of-aerial-imagery}
\showURL{%
Retrieved October 18, 2024 from \tempurl}


\bibitem[exa(2023)]%
        {exakratos}
Cubesat Market \bibinfo{year}{2023}\natexlab{}.
\newblock \bibinfo{booktitle}{\emph{KRATOS 1U Ready to Fly Cubesat Platform}}.
\newblock Cubesat Market.
\newblock
\urldef\tempurl%
\url{https://www.cubesat.market/kratos1uplatform}
\showURL{%
Retrieved August 28, 2024 from \tempurl}


\bibitem[sen(2023)]%
        {sentinel2}
European Space Agency \bibinfo{year}{2023}\natexlab{}.
\newblock \bibinfo{booktitle}{\emph{Overview of Sentinel-2 Mission}}.
\newblock European Space Agency.
\newblock
\urldef\tempurl%
\url{https://sentiwiki.copernicus.eu/web/s2-mission}
\showURL{%
Retrieved October 18, 2024 from \tempurl}


\bibitem[pla(2023)]%
        {planet2023product}
Planet Labs \bibinfo{year}{2023}\natexlab{}.
\newblock \bibinfo{booktitle}{\emph{Planet Imagery Product Specifications}}.
\newblock Planet Labs.
\newblock
\urldef\tempurl%
\url{https://assets.planet.com/docs/Planet_Combined_Imagery_Product_Specs_letter_screen.pdf}
\showURL{%
Retrieved April 9, 2025 from \tempurl}


\bibitem[spa(2023)]%
        {spacecloud}
Unibap AB \bibinfo{year}{2023}\natexlab{}.
\newblock \bibinfo{booktitle}{\emph{SpaceCloudOS \& SpaceCloudFW}}.
\newblock Unibap AB.
\newblock
\urldef\tempurl%
\url{https://unibap.com/wp-content/uploads/2023/12/spacecloud-os-fw.pdf}
\showURL{%
Retrieved October 18, 2024 from \tempurl}


\bibitem[ucs(2023)]%
        {ucssd}
Union of Concerned Scientists \bibinfo{year}{2023}\natexlab{}.
\newblock \bibinfo{booktitle}{\emph{UCS Satellite Database}}.
\newblock Union of Concerned Scientists.
\newblock
\urldef\tempurl%
\url{https://www.ucsusa.org/resources/satellite-database}
\showURL{%
Retrieved August 23, 2024 from \tempurl}


\bibitem[vii(2023)]%
        {viirs}
National Environmental Satellite, Data, and Information Service (NESDIS)
  \bibinfo{year}{2023}\natexlab{}.
\newblock \bibinfo{booktitle}{\emph{Visible Infrared Imaging Radiometer Suite
  (VIIRS)}}.
\newblock National Environmental Satellite, Data, and Information Service
  (NESDIS).
\newblock
\urldef\tempurl%
\url{https://www.nesdis.noaa.gov/our-satellites/currently-flying/joint-polar-satellite-system/visible-infrared-imaging-radiometer-suite-viirs}
\showURL{%
Retrieved October 18, 2024 from \tempurl}


\bibitem[orb(2024)]%
        {orbitsedge}
OrbitsEdge, Inc. \bibinfo{year}{2024}\natexlab{}.
\newblock \bibinfo{booktitle}{\emph{Edge Computing \& Micro Datacenters in
  Space}}.
\newblock OrbitsEdge, Inc.
\newblock
\urldef\tempurl%
\url{https://orbitsedge.com/edge-in-space}
\showURL{%
Retrieved October 18, 2024 from \tempurl}


\bibitem[pum(2025)]%
        {pumpkin}
Pumpkin Space Systems \bibinfo{year}{2025}\natexlab{}.
\newblock \bibinfo{booktitle}{\emph{135W Dual Articulated Deployable Solar
  Array}}.
\newblock Pumpkin Space Systems.
\newblock
\urldef\tempurl%
\url{https://www.pumpkinspace.com/store/p215/135W_Dual_Articulated_Deployable_Solar_Array.html}
\showURL{%
Retrieved April 9, 2025 from \tempurl}


\bibitem[nas(2025)]%
        {nasa2025cutting}
NASA Spinoff Technology Transfer Program \bibinfo{year}{2025}\natexlab{}.
\newblock \bibinfo{booktitle}{\emph{Cutting-Edge Computing Goes Spaceborne}}.
\newblock NASA Spinoff Technology Transfer Program.
\newblock
\urldef\tempurl%
\url{https://spinoff.nasa.gov/Cutting-Edge_Computing_Goes_Spaceborne}
\showURL{%
Retrieved April 9, 2025 from \tempurl}


\bibitem[Aaba(2021)]%
        {wildfire}
\bibfield{author}{\bibinfo{person}{Abdelghani Aaba}.}
  \bibinfo{year}{2021}\natexlab{}.
\newblock \bibinfo{booktitle}{\emph{Wildfire Prediction Dataset (Satellite
  Images)}}.
\newblock
\urldef\tempurl%
\url{https://www.kaggle.com/datasets/abdelghaniaaba/wildfire-prediction-dataset}
\showURL{%
Retrieved October 18, 2024 from \tempurl}


\bibitem[Agache et~al\mbox{.}(2020)]%
        {agache2020firecracker}
\bibfield{author}{\bibinfo{person}{Alexandru Agache}, \bibinfo{person}{Marc
  Brooker}, \bibinfo{person}{Alexandra Iordache}, \bibinfo{person}{Anthony
  Liguori}, \bibinfo{person}{Rolf Neugebauer}, \bibinfo{person}{Phil Piwonka},
  {and} \bibinfo{person}{Diana-Maria Popa}.} \bibinfo{year}{2020}\natexlab{}.
\newblock \showarticletitle{Firecracker: Lightweight virtualization for
  serverless applications}. In \bibinfo{booktitle}{\emph{Proceedings of the
  17th USENIX Symposium on Networked Systems Design and Implementation}} (Santa
  Clara, CA, USA) \emph{(\bibinfo{series}{NSDI '20})}.
  \bibinfo{publisher}{USENIX Association}, \bibinfo{address}{Berkeley, CA,
  USA}, \bibinfo{pages}{419--434}.
\newblock


\bibitem[Antonio-Javier~Gallego and Gil(2018)]%
        {gallego2018automatic}
\bibfield{author}{\bibinfo{person}{Antonio~Pertusa Antonio-Javier~Gallego}
  {and} \bibinfo{person}{Pablo Gil}.} \bibinfo{year}{2018}\natexlab{}.
\newblock \showarticletitle{Automatic Ship Classification from Optical Aerial
  Images with Convolutional Neural Networks}.
\newblock \bibinfo{journal}{\emph{Remote Sensing}} \bibinfo{volume}{10},
  \bibinfo{number}{4}, Article \bibinfo{articleno}{511} (\bibinfo{date}{April}
  \bibinfo{year}{2018}), \bibinfo{numpages}{20}~pages.
\newblock
\showISSN{2072-4292}
\href{https://doi.org/10.3390/rs10040511}{doi:\nolinkurl{10.3390/rs10040511}}


\bibitem[Arechiga et~al\mbox{.}(2018)]%
        {arechiga2018onboard}
\bibfield{author}{\bibinfo{person}{Austin~P. Arechiga},
  \bibinfo{person}{Alan~J. Michaels}, {and} \bibinfo{person}{Jonathan~T.
  Black}.} \bibinfo{year}{2018}\natexlab{}.
\newblock \showarticletitle{Onboard Image Processing for Small Satellites}. In
  \bibinfo{booktitle}{\emph{Proceedings of the IEEE National Aerospace and
  Electronics Conference}} (Dayton, OH, USA) \emph{(\bibinfo{series}{NAECON
  2018})}. \bibinfo{publisher}{IEEE}, \bibinfo{address}{New York, NY, USA},
  \bibinfo{pages}{234--240}.
\newblock
\href{https://doi.org/10.1109/NAECON.2018.8556744}{doi:\nolinkurl{10.1109/NAECON.2018.8556744}}


\bibitem[Aslanpour et~al\mbox{.}(2021)]%
        {aslanpour2021serverless}
\bibfield{author}{\bibinfo{person}{Mohammad~S. Aslanpour},
  \bibinfo{person}{Adel~N. Toosi}, \bibinfo{person}{Claudio Cicconetti},
  \bibinfo{person}{Bahman Javadi}, \bibinfo{person}{Peter Sbarski},
  \bibinfo{person}{Davide Taibi}, \bibinfo{person}{Marcos Assuncao},
  \bibinfo{person}{Sukhpal~Singh Gill}, \bibinfo{person}{Raj Gaire}, {and}
  \bibinfo{person}{Schahram Dustdar}.} \bibinfo{year}{2021}\natexlab{}.
\newblock \showarticletitle{Serverless Edge Computing: Vision and Challenges}.
  In \bibinfo{booktitle}{\emph{Proceedings of the 2021 Australasian Computer
  Science Week Multiconference}} (Dunedin, New Zealand)
  \emph{(\bibinfo{series}{ACSW '21})}. \bibinfo{publisher}{Association for
  Computing Machinery}, \bibinfo{address}{New York, NY, USA},
  \bibinfo{pages}{1--10}.
\newblock
\href{https://doi.org/10.1145/3437378.3444367}{doi:\nolinkurl{10.1145/3437378.3444367}}


\bibitem[Bati\v{c}(2018)]%
        {s2cloudless}
\bibfield{author}{\bibinfo{person}{Matej Bati\v{c}}.}
  \bibinfo{year}{2018}\natexlab{}.
\newblock \bibinfo{booktitle}{\emph{Sentinel Hub Cloud Detector —
  s2cloudless}}.
\newblock Sentinel Hub.
\newblock
\urldef\tempurl%
\url{https://medium.com/sentinel-hub/sentinel-hub-cloud-detector-s2cloudless-a67d263d3025}
\showURL{%
Retrieved October 18, 2024 from \tempurl}


\bibitem[Boley and Byers(2021)]%
        {boley2021satellite}
\bibfield{author}{\bibinfo{person}{Aaron~C. Boley} {and}
  \bibinfo{person}{Michael Byers}.} \bibinfo{year}{2021}\natexlab{}.
\newblock \showarticletitle{Satellite mega-constellations create risks in Low
  Earth Orbit, the atmosphere and on Earth}.
\newblock \bibinfo{journal}{\emph{Scientific Reports}}  \bibinfo{volume}{11},
  Article \bibinfo{articleno}{10642} (\bibinfo{date}{May}
  \bibinfo{year}{2021}), \bibinfo{numpages}{8}~pages.
\newblock
\showISSN{2045-2322}
\href{https://doi.org/10.1038/s41598-021-89909-7}{doi:\nolinkurl{10.1038/s41598-021-89909-7}}


\bibitem[Bouwmeester and Guo(2010)]%
        {bouwmeester2010survey}
\bibfield{author}{\bibinfo{person}{Jasper Bouwmeester} {and}
  \bibinfo{person}{Jian Guo}.} \bibinfo{year}{2010}\natexlab{}.
\newblock \showarticletitle{Survey of worldwide pico- and nanosatellite
  missions, distributions and subsystem technology}.
\newblock \bibinfo{journal}{\emph{Acta Astronautica}} \bibinfo{volume}{67},
  \bibinfo{number}{7--8} (\bibinfo{date}{June} \bibinfo{year}{2010}),
  \bibinfo{pages}{854--862}.
\newblock
\showISSN{0094-5765}
\href{https://doi.org/10.1016/j.actaastro.2010.06.004}{doi:\nolinkurl{10.1016/j.actaastro.2010.06.004}}


\bibitem[Braaten et~al\mbox{.}(2015)]%
        {braaten2015automated}
\bibfield{author}{\bibinfo{person}{Justin~D. Braaten},
  \bibinfo{person}{Warren~B. Cohen}, {and} \bibinfo{person}{Zhiqiang Yang}.}
  \bibinfo{year}{2015}\natexlab{}.
\newblock \showarticletitle{Automated cloud and cloud shadow identification in
  Landsat MSS imagery for temperate ecosystems}.
\newblock \bibinfo{journal}{\emph{Remote Sensing of Environment}}
  \bibinfo{volume}{169} (\bibinfo{date}{Aug.} \bibinfo{year}{2015}),
  \bibinfo{pages}{128--138}.
\newblock
\showISSN{0034--4257}
\href{https://doi.org/10.1016/j.rse.2015.08.006}{doi:\nolinkurl{10.1016/j.rse.2015.08.006}}


\bibitem[Brenner and Kapitza(2019)]%
        {brenner2019trust}
\bibfield{author}{\bibinfo{person}{Stefan Brenner} {and}
  \bibinfo{person}{R{\"u}diger Kapitza}.} \bibinfo{year}{2019}\natexlab{}.
\newblock \showarticletitle{Trust more, serverless}. In
  \bibinfo{booktitle}{\emph{Proceedings of the 12th ACM International
  Conference on Systems and Storage}} (Haifa, Israel)
  \emph{(\bibinfo{series}{SYSTOR '19})}. \bibinfo{publisher}{Association for
  Computing Machinery}, \bibinfo{address}{New York, NY, USA},
  \bibinfo{pages}{33--43}.
\newblock
\href{https://doi.org/10.1145/3319647.3325825}{doi:\nolinkurl{10.1145/3319647.3325825}}


\bibitem[Cao et~al\mbox{.}(2013)]%
        {viirsreport}
\bibfield{author}{\bibinfo{person}{Changyong Cao},
  \bibinfo{person}{Xiaoxiong~(Jack) Xiong}, \bibinfo{person}{Robert Wolfe},
  \bibinfo{person}{Frank DeLuccia}, \bibinfo{person}{Quanhua~(Mark) Liu},
  \bibinfo{person}{Slawomir Blonski}, \bibinfo{person}{Guoqing~(Gary) Lin},
  \bibinfo{person}{Masahiro Nishihama}, \bibinfo{person}{Dave Pogorzala},
  \bibinfo{person}{Hassan Oudrari}, {and} \bibinfo{person}{Don Hillger}.}
  \bibinfo{year}{2013}\natexlab{}.
\newblock \bibinfo{booktitle}{\emph{Visible Infrared Imaging Radiometer Suite
  (VIIRS) Sensor Data Record (SDR) User's Guide}}.
\newblock \bibinfo{type}{{T}echnical {R}eport} 142.
  \bibinfo{institution}{National Environmental Satellite, Data, and Information
  Service, National Oceanic and Atmospheric Administration, U.S. DEPARTMENT OF
  COMMERCE}, \bibinfo{address}{Washington, D.C., USA}.
\newblock
\urldef\tempurl%
\url{https://nsidc.org/sites/default/files/viirs-sdr-users-guide.pdf}
\showURL{%
\tempurl}


\bibitem[Cappaert(2018)]%
        {cappaert2018building}
\bibfield{author}{\bibinfo{person}{Jeroen Cappaert}.}
  \bibinfo{year}{2018}\natexlab{}.
\newblock \showarticletitle{Building, Deploying and Operating a Cubesat
  Constellation - Exploring the Less Obvious Reasons Space is Hard}. In
  \bibinfo{booktitle}{\emph{Proceedings of the 32nd Annual Small Satellite
  Conference}} (Logan, UT, USA) \emph{(\bibinfo{series}{SmallSat '18})}.
  \bibinfo{publisher}{Utah State University}, \bibinfo{address}{Logan, UT,
  USA}.
\newblock


\bibitem[Carnahan et~al\mbox{.}(2022)]%
        {mehrparvar2022cubesat}
\bibfield{author}{\bibinfo{person}{Justin Carnahan}, \bibinfo{person}{Amy
  Hutputanasin}, \bibinfo{person}{Alicia Johnstone}, \bibinfo{person}{Wenschel
  Lan}, \bibinfo{person}{Simon Lee}, \bibinfo{person}{Arash Mehrpavar},
  \bibinfo{person}{Riki Munakata}, \bibinfo{person}{David Pignatelli}, {and}
  \bibinfo{person}{Armen Toorian}.} \bibinfo{year}{2022}\natexlab{}.
\newblock \bibinfo{booktitle}{\emph{Cubesat Design Specification (1U -- 12U),
  Rev. 14.1}}.
\newblock \bibinfo{type}{{T}echnical {R}eport} CP-CDS-R14.1.
  \bibinfo{institution}{California Polytechnic State University},
  \bibinfo{address}{San Luis Obispo, CA, USA}.
\newblock
\urldef\tempurl%
\url{https://www.cubesat.org/s/CDS-REV14_1-2022-02-09.pdf}
\showURL{%
\tempurl}


\bibitem[Chen et~al\mbox{.}(2017)]%
        {chen2017heavy}
\bibfield{author}{\bibinfo{person}{Dakai Chen}, \bibinfo{person}{Edward
  Wilcox}, \bibinfo{person}{Raymond~L. Ladbury}, \bibinfo{person}{Christina
  Seidleck}, \bibinfo{person}{Hak Kim}, \bibinfo{person}{Anthony Phan}, {and}
  \bibinfo{person}{Kenneth~A. LaBel}.} \bibinfo{year}{2017}\natexlab{}.
\newblock \showarticletitle{Heavy Ion and Proton-Induced Single Event Upset
  Characteristics of a 3-D NAND Flash Memory}.
\newblock \bibinfo{journal}{\emph{IEEE Transactions on Nuclear Science}}
  \bibinfo{volume}{65}, \bibinfo{number}{1} (\bibinfo{date}{Oct.}
  \bibinfo{year}{2017}), \bibinfo{pages}{19--26}.
\newblock
\showISSN{0018-9499}
\href{https://doi.org/10.1109/TNS.2017.2764852}{doi:\nolinkurl{10.1109/TNS.2017.2764852}}


\bibitem[Damkalis et~al\mbox{.}(2024)]%
        {kanyinidb}
\bibfield{author}{\bibinfo{person}{Alfredos-Panagiotis Damkalis},
  \bibinfo{person}{Daniel Esparon}, {and} \bibinfo{person}{Jack Philpott}.}
  \bibinfo{year}{2024}\natexlab{}.
\newblock \bibinfo{booktitle}{\emph{SatNOGS DB -- Kanyini}}.
\newblock Libre Space Foundation.
\newblock
\urldef\tempurl%
\url{https://db.satnogs.org/satellite/98890#data}
\showURL{%
Retrieved August 29, 2024 from \tempurl}


\bibitem[Denby et~al\mbox{.}(2023)]%
        {denby2023kodan}
\bibfield{author}{\bibinfo{person}{Bradley Denby}, \bibinfo{person}{Krishna
  Chintalapudi}, \bibinfo{person}{Ranveer Chandra}, \bibinfo{person}{Brandon
  Lucia}, {and} \bibinfo{person}{Shadi Noghabi}.}
  \bibinfo{year}{2023}\natexlab{}.
\newblock \showarticletitle{Kodan: Addressing the Computational Bottleneck in
  Space}. In \bibinfo{booktitle}{\emph{Proceedings of the 28th ACM
  International Conference on Architectural Support for Programming Languages
  and Operating Systems}} (Vancouver, BC, Canada)
  \emph{(\bibinfo{series}{ASPLOS '23})}. \bibinfo{publisher}{Association for
  Computing Machinery}, \bibinfo{address}{New York, NY, USA},
  \bibinfo{pages}{392--403}.
\newblock
\href{https://doi.org/10.1145/3582016.3582043}{doi:\nolinkurl{10.1145/3582016.3582043}}


\bibitem[Denby and Lucia(2020)]%
        {denby2020orbital}
\bibfield{author}{\bibinfo{person}{Bradley Denby} {and}
  \bibinfo{person}{Brandon Lucia}.} \bibinfo{year}{2020}\natexlab{}.
\newblock \showarticletitle{Orbital Edge Computing: Nanosatellite
  Constellations as a New Class of Computer System}. In
  \bibinfo{booktitle}{\emph{Proceedings of the Twenty-Fifth International
  Conference on Architectural Support for Programming Languages and Operating
  Systems}} (Lausanne, Switzerland) \emph{(\bibinfo{series}{ASPLOS '20})}.
  \bibinfo{publisher}{Association for Computing Machinery},
  \bibinfo{address}{New York, NY, USA}, \bibinfo{pages}{939--954}.
\newblock
\href{https://doi.org/10.1145/3373376.3378473}{doi:\nolinkurl{10.1145/3373376.3378473}}


\bibitem[Devaraj et~al\mbox{.}(2019)]%
        {devaraj2019planet}
\bibfield{author}{\bibinfo{person}{Kiruthika Devaraj}, \bibinfo{person}{Matt
  Ligon}, \bibinfo{person}{Eric Blossom}, \bibinfo{person}{Joseph Breu},
  \bibinfo{person}{Bryan Klofas}, \bibinfo{person}{Kyle Colton}, {and}
  \bibinfo{person}{Ryan Kingsbury}.} \bibinfo{year}{2019}\natexlab{}.
\newblock \showarticletitle{Planet high speed radio: Crossing Gbps from a 3U
  cubesat}. In \bibinfo{booktitle}{\emph{Proceedings of the 33rd Annual Small
  Satellite Conference}} (Logan, UT, USA) \emph{(\bibinfo{series}{SmallSat
  '19})}. \bibinfo{publisher}{Utah State University}, \bibinfo{address}{Logan,
  UT, USA}.
\newblock


\bibitem[Drusch et~al\mbox{.}(2012)]%
        {drusch2012sentinel}
\bibfield{author}{\bibinfo{person}{Matthias Drusch}, \bibinfo{person}{Umberto
  Del~Bello}, \bibinfo{person}{S{\'e}bastien Carlier}, \bibinfo{person}{Olivier
  Colin}, \bibinfo{person}{Veronica Fernandez}, \bibinfo{person}{Ferran
  Gascon}, \bibinfo{person}{Bianca Hoersch}, \bibinfo{person}{Claudia Isola},
  \bibinfo{person}{Paolo Laberinti}, \bibinfo{person}{Philippe Martimort},
  {et~al\mbox{.}}} \bibinfo{year}{2012}\natexlab{}.
\newblock \showarticletitle{Sentinel-2: ESA's optical high-resolution mission
  for GMES operational services}.
\newblock \bibinfo{journal}{\emph{Remote sensing of Environment}}
  \bibinfo{volume}{120} (\bibinfo{date}{Feb.} \bibinfo{year}{2012}),
  \bibinfo{pages}{25--36}.
\newblock
\showISSN{0034--4257}
\href{https://doi.org/10.1016/j.rse.2011.11.026}{doi:\nolinkurl{10.1016/j.rse.2011.11.026}}


\bibitem[Elgamal et~al\mbox{.}(2018)]%
        {elgamal2018costless}
\bibfield{author}{\bibinfo{person}{Tarek Elgamal}, \bibinfo{person}{Atul
  Sandur}, \bibinfo{person}{Klara Nahrstedt}, {and} \bibinfo{person}{Gul
  Agha}.} \bibinfo{year}{2018}\natexlab{}.
\newblock \showarticletitle{Costless: Optimizing Cost of Serverless Computing
  through Function Fusion and Placement}. In
  \bibinfo{booktitle}{\emph{Proceedings of the 2018 IEEE/ACM Symposium on Edge
  Computing}} (Seattle, WA, USA) \emph{(\bibinfo{series}{SEC '18})}.
  \bibinfo{publisher}{IEEE}, \bibinfo{address}{New York, NY, USA},
  \bibinfo{pages}{300--312}.
\newblock
\href{https://doi.org/10.1109/SEC.2018.00029}{doi:\nolinkurl{10.1109/SEC.2018.00029}}


\bibitem[Ferreira et~al\mbox{.}(2024)]%
        {ferreira2024potential}
\bibfield{author}{\bibinfo{person}{Jos\'e~P. Ferreira}, \bibinfo{person}{Ziyu
  Huang}, \bibinfo{person}{Ken-ichi Nomura}, {and} \bibinfo{person}{Joseph
  Wang}.} \bibinfo{year}{2024}\natexlab{}.
\newblock \showarticletitle{Potential Ozone Depletion From Satellite Demise
  During Atmospheric Reentry in the Era of Mega-Constellations}.
\newblock \bibinfo{journal}{\emph{Geophysical Research Letters}}
  \bibinfo{volume}{51}, \bibinfo{number}{11}, Article
  \bibinfo{articleno}{e2024GL109280} (\bibinfo{date}{June}
  \bibinfo{year}{2024}).
\newblock
\showISSN{0094-8276}
\href{https://doi.org/10.1029/2024GL109280}{doi:\nolinkurl{10.1029/2024GL109280}}


\bibitem[Furano et~al\mbox{.}(2020)]%
        {furano2020towards}
\bibfield{author}{\bibinfo{person}{Gianluca Furano}, \bibinfo{person}{Gabriele
  Meoni}, \bibinfo{person}{Aubrey Dunne}, \bibinfo{person}{David Moloney},
  \bibinfo{person}{Veronique Ferlet-Cavrois}, \bibinfo{person}{Antonis
  Tavoularis}, \bibinfo{person}{Jonathan Byrne}, \bibinfo{person}{L{\'e}onie
  Buckley}, \bibinfo{person}{Mihalis Psarakis}, \bibinfo{person}{Kay-Obbe
  Voss}, {and} \bibinfo{person}{Luca Fanucci}.}
  \bibinfo{year}{2020}\natexlab{}.
\newblock \showarticletitle{Towards the Use of Artificial Intelligence on the
  Edge in Space Systems: Challenges and Opportunities}.
\newblock \bibinfo{journal}{\emph{IEEE Aerospace and Electronic Systems
  Magazine}} \bibinfo{volume}{35}, \bibinfo{number}{12} (\bibinfo{date}{Dec.}
  \bibinfo{year}{2020}), \bibinfo{pages}{44--56}.
\newblock
\showISSN{0885-8985}
\href{https://doi.org/10.1109/MAES.2020.3008468}{doi:\nolinkurl{10.1109/MAES.2020.3008468}}


\bibitem[Furutanpey et~al\mbox{.}(2025)]%
        {furutanpey2024fool}
\bibfield{author}{\bibinfo{person}{Alireza Furutanpey}, \bibinfo{person}{Qiyang
  Zhang}, \bibinfo{person}{Philipp Raith}, \bibinfo{person}{Tobias
  Pfandzelter}, \bibinfo{person}{Shangguang Wang}, {and}
  \bibinfo{person}{Schahram Dustdar}.} \bibinfo{year}{2025}\natexlab{}.
\newblock \showarticletitle{FOOL: Addressing the Downlink Bottleneck in
  Satellite Computing with Neural Feature Compression}.
\newblock \bibinfo{journal}{\emph{IEEE Transactions on Mobile Computing}}
  (\bibinfo{date}{Feb.} \bibinfo{year}{2025}).
\newblock
\showISSN{1536-1233}
\href{https://doi.org/10.1109/TMC.2025.3544516}{doi:\nolinkurl{10.1109/TMC.2025.3544516}}


\bibitem[Garcia(2021)]%
        {garcia2021electric}
\bibfield{author}{\bibinfo{person}{Jose~L. Garcia}.}
  \bibinfo{year}{2021}\natexlab{}.
\newblock \showarticletitle{9 - Electric power systems}.
\newblock In \bibinfo{booktitle}{\emph{Cubesat Handbook}},
  \bibfield{editor}{\bibinfo{person}{Chantal Cappelletti},
  \bibinfo{person}{Simone Battistini}, {and} \bibinfo{person}{Benjamin~K.
  Malphrus}} (Eds.). \bibinfo{publisher}{Academic Press},
  \bibinfo{pages}{185--197}.
\newblock
\showISBNx{978-0-12-817884-3}
\href{https://doi.org/10.1016/B978-0-12-817884-3.00009-6}{doi:\nolinkurl{10.1016/B978-0-12-817884-3.00009-6}}


\bibitem[G{\'e}hberger and Kov{\'a}cs(2022)]%
        {gehberger2022cooling}
\bibfield{author}{\bibinfo{person}{D{\'a}niel G{\'e}hberger} {and}
  \bibinfo{person}{D{\'a}vid Kov{\'a}cs}.} \bibinfo{year}{2022}\natexlab{}.
\newblock \showarticletitle{Cooling down faas: Towards getting rid of warm
  starts}.
\newblock  (\bibinfo{date}{June} \bibinfo{year}{2022}).
\newblock
\showeprint{2206.00599}


\bibitem[Giuffrida et~al\mbox{.}(2020)]%
        {giuffrida2020cloudscout}
\bibfield{author}{\bibinfo{person}{Gianluca Giuffrida},
  \bibinfo{person}{Lorenzo Diana}, \bibinfo{person}{Francesco de Gioia},
  \bibinfo{person}{Gionata Benelli}, \bibinfo{person}{Gabriele Meoni},
  \bibinfo{person}{Massimiliano Donati}, {and} \bibinfo{person}{Luca Fanucci}.}
  \bibinfo{year}{2020}\natexlab{}.
\newblock \showarticletitle{CloudScout: A deep neural network for on-board
  cloud detection on hyperspectral images}.
\newblock \bibinfo{journal}{\emph{Remote Sensing}} \bibinfo{volume}{12},
  \bibinfo{number}{14}, Article \bibinfo{articleno}{2205} (\bibinfo{date}{July}
  \bibinfo{year}{2020}), \bibinfo{numpages}{17}~pages.
\newblock
\showISSN{2072-4292}
\href{https://doi.org/10.3390/rs12142205}{doi:\nolinkurl{10.3390/rs12142205}}


\bibitem[Giuffrida et~al\mbox{.}(2021)]%
        {giuffrida2021varphi}
\bibfield{author}{\bibinfo{person}{Gianluca Giuffrida}, \bibinfo{person}{Luca
  Fanucci}, \bibinfo{person}{Gabriele Meoni}, \bibinfo{person}{Matej
  Bati{\v{c}}}, \bibinfo{person}{L{\'e}onie Buckley}, \bibinfo{person}{Aubrey
  Dunne}, \bibinfo{person}{Chris Van~Dijk}, \bibinfo{person}{Marco Esposito},
  \bibinfo{person}{John Hefele}, \bibinfo{person}{Gianluca Vercruyssen,
  Nathan~Furano}, \bibinfo{person}{Massimiliano Pastena}, {and}
  \bibinfo{person}{Josef Aschbacher}.} \bibinfo{year}{2021}\natexlab{}.
\newblock \showarticletitle{The $\Phi$-Sat-1 Mission: The First On-Board Deep
  Neural Network Demonstrator for Satellite Earth Observation}.
\newblock \bibinfo{journal}{\emph{IEEE Transactions on Geoscience and Remote
  Sensing}}  \bibinfo{volume}{60} (\bibinfo{date}{Nov.} \bibinfo{year}{2021}),
  \bibinfo{pages}{1--14}.
\newblock
\showISSN{0196-2892}
\href{https://doi.org/10.1109/TGRS.2021.3125567}{doi:\nolinkurl{10.1109/TGRS.2021.3125567}}


\bibitem[Greene et~al\mbox{.}(2022)]%
        {greene2023system}
\bibfield{author}{\bibinfo{person}{Jared~Michael Greene},
  \bibinfo{person}{Mohammed~Faraz Admani}, \bibinfo{person}{Jacob~Nelson
  Glueck}, \bibinfo{person}{Sergii Ziuzin}, \bibinfo{person}{Francesco
  De~Paolis}, \bibinfo{person}{Dhruv Dawar}, {and} \bibinfo{person}{Christopher
  Yu}.} \bibinfo{year}{2022}\natexlab{}.
\newblock \bibinfo{title}{System and method of providing access to compute
  resources distributed across a group of satellites}.
\newblock
\newblock
\shownote{US Patent App. US20230164089A1: filed Sep. 28., 2022}.


\bibitem[Hegazi et~al\mbox{.}(2023)]%
        {hegazi2023prediction}
\bibfield{author}{\bibinfo{person}{Ehab~H. Hegazi},
  \bibinfo{person}{Abdellateif~A. Samak}, \bibinfo{person}{Lingbo Yang},
  \bibinfo{person}{Ran Huang}, {and} \bibinfo{person}{Jingfeng Huang}.}
  \bibinfo{year}{2023}\natexlab{}.
\newblock \showarticletitle{Prediction of Soil Moisture Content from Sentinel-2
  Images Using Convolutional Neural Network (CNN)}.
\newblock \bibinfo{journal}{\emph{Agronomy}} \bibinfo{volume}{13},
  \bibinfo{number}{3}, Article \bibinfo{articleno}{656} (\bibinfo{date}{Feb.}
  \bibinfo{year}{2023}), \bibinfo{numpages}{18}~pages.
\newblock
\showISSN{2073--4395}
\href{https://doi.org/10.3390/rs10040511}{doi:\nolinkurl{10.3390/rs10040511}}


\bibitem[Helber et~al\mbox{.}(2019)]%
        {helber2019eurosat}
\bibfield{author}{\bibinfo{person}{Patrick Helber}, \bibinfo{person}{Benjamin
  Bischke}, \bibinfo{person}{Andreas Dengel}, {and} \bibinfo{person}{Damian
  Borth}.} \bibinfo{year}{2019}\natexlab{}.
\newblock \showarticletitle{EuroSAT: A Novel Dataset and Deep Learning
  Benchmark for Land Use and Land Cover Classification}.
\newblock \bibinfo{journal}{\emph{IEEE Journal of Selected Topics in Applied
  Earth Observations and Remote Sensing}} \bibinfo{volume}{12},
  \bibinfo{number}{7} (\bibinfo{date}{June} \bibinfo{year}{2019}),
  \bibinfo{pages}{2217--2226}.
\newblock
\showISSN{1939-1404}
\href{https://doi.org/10.1109/JSTARS.2019.2918242}{doi:\nolinkurl{10.1109/JSTARS.2019.2918242}}


\bibitem[Hendrickson et~al\mbox{.}(2016)]%
        {hendrickson2016serverless}
\bibfield{author}{\bibinfo{person}{Scott Hendrickson}, \bibinfo{person}{Stephen
  Sturdevant}, \bibinfo{person}{Tyler Harter}, \bibinfo{person}{Venkateshwaran
  Venkataramani}, \bibinfo{person}{Andrea~C. Arpaci-Dusseau}, {and}
  \bibinfo{person}{Remzi~H. Arpaci-Dusseau}.} \bibinfo{year}{2016}\natexlab{}.
\newblock \showarticletitle{Serverless Computation with OpenLambda}. In
  \bibinfo{booktitle}{\emph{Proceedings of the 8th USENIX Workshop on Hot
  Topics in Cloud Computing}} (Denver, CO, USA)
  \emph{(\bibinfo{series}{HotCloud '16})}. \bibinfo{publisher}{USENIX
  Association}, \bibinfo{address}{Berkeley, CA, USA}, \bibinfo{pages}{33--39}.
\newblock


\bibitem[Henkel et~al\mbox{.}(2024)]%
        {henkel2023mitigating}
\bibfield{author}{\bibinfo{person}{Maximilian Henkel},
  \bibinfo{person}{Vladimir Zelenevskiy}, \bibinfo{person}{Georges Labrèche},
  \bibinfo{person}{Rodrigo Laurinovics}, \bibinfo{person}{David Evans},
  \bibinfo{person}{Dominik Marszk}, {and} \bibinfo{person}{Omiros~Papadatos
  Vasilakis}.} \bibinfo{year}{2024}\natexlab{}.
\newblock \showarticletitle{Mitigating and Recovering from Radiation Induced
  Faults in Non-Hardened Spacecraft Flash Memory}. In
  \bibinfo{booktitle}{\emph{Proceedings of the 2024 IEEE Aerospace Conference}}
  (Big Sky, MT, USA) \emph{(\bibinfo{series}{AERO '24})}.
  \bibinfo{publisher}{IEEE}, \bibinfo{address}{New York, NY, USA},
  \bibinfo{pages}{1--8}.
\newblock
\href{https://doi.org/10.1109/AERO58975.2024.10521125}{doi:\nolinkurl{10.1109/AERO58975.2024.10521125}}


\bibitem[Horine(2021)]%
        {horine2021creating}
\bibfield{author}{\bibinfo{person}{Brent Horine}.}
  \bibinfo{year}{2021}\natexlab{}.
\newblock \showarticletitle{Creating a Marketplace for a Constellation as a
  Service}. In \bibinfo{booktitle}{\emph{Proceedings of the 38th Annual Small
  Satellite Conference}} (Logan, UT, USA) \emph{(\bibinfo{series}{SmallSat
  '21})}. \bibinfo{publisher}{Utah State University}, \bibinfo{address}{Logan,
  UT, USA}.
\newblock


\bibitem[Jonas et~al\mbox{.}(2019)]%
        {jonas2019cloud}
\bibfield{author}{\bibinfo{person}{Eric Jonas}, \bibinfo{person}{Johann
  Schleier-Smith}, \bibinfo{person}{Vikram Sreekanti},
  \bibinfo{person}{Chia-Che Tsai}, \bibinfo{person}{Anurag Khandelwal},
  \bibinfo{person}{Qifan Pu}, \bibinfo{person}{Vaishaal Shankar},
  \bibinfo{person}{Joao Menezes~Carreira}, \bibinfo{person}{Karl Krauth},
  \bibinfo{person}{Neeraja Yadwadkar}, \bibinfo{person}{Joseph Gonzalez},
  \bibinfo{person}{Raluca~Ada Popa}, \bibinfo{person}{Ion Stoica}, {and}
  \bibinfo{person}{David~A. Patterson}.} \bibinfo{year}{2019}\natexlab{}.
\newblock \bibinfo{booktitle}{\emph{Cloud Programming Simplified: A Berkeley
  View on Serverless Computing}}.
\newblock \bibinfo{type}{{T}echnical {R}eport} UCB/EECS-2019-3.
  \bibinfo{institution}{EECS Department, University of California, Berkeley},
  \bibinfo{address}{Berkeley, CA, USA}.
\newblock
\urldef\tempurl%
\url{https://www2.eecs.berkeley.edu/Pubs/TechRpts/2019/EECS-2019-3.html}
\showURL{%
\tempurl}


\bibitem[Kessler and Cour-Palais(1978)]%
        {kessler1978collision}
\bibfield{author}{\bibinfo{person}{Donald~J. Kessler} {and}
  \bibinfo{person}{Burton~G. Cour-Palais}.} \bibinfo{year}{1978}\natexlab{}.
\newblock \showarticletitle{Collision frequency of artificial satellites: The
  creation of a debris belt}.
\newblock \bibinfo{journal}{\emph{Journal of Geophysical Research: Space
  Physics}} \bibinfo{volume}{83}, \bibinfo{number}{6} (\bibinfo{date}{June}
  \bibinfo{year}{1978}), \bibinfo{pages}{2637--2646}.
\newblock
\showISSN{2169-9380}
\href{https://doi.org/10.1029/JA083iA06p02637}{doi:\nolinkurl{10.1029/JA083iA06p02637}}


\bibitem[King et~al\mbox{.}(2013)]%
        {king2013spatial}
\bibfield{author}{\bibinfo{person}{Michael~D. King}, \bibinfo{person}{Steven
  Platnick}, \bibinfo{person}{W.~Paul Menzel}, \bibinfo{person}{Steven~A.
  Ackerman}, {and} \bibinfo{person}{Paul~A. Hubanks}.}
  \bibinfo{year}{2013}\natexlab{}.
\newblock \showarticletitle{Spatial and Temporal Distribution of Clouds
  Observed by MODIS Onboard the Terra and Aqua Satellites}.
\newblock \bibinfo{journal}{\emph{IEEE Transactions on Geoscience and Remote
  Sensing}} \bibinfo{volume}{51}, \bibinfo{number}{7} (\bibinfo{date}{Jan.}
  \bibinfo{year}{2013}), \bibinfo{pages}{3826--3852}.
\newblock
\showISSN{0196--2892}
\href{https://doi.org/10.1109/TGRS.2012.2227333}{doi:\nolinkurl{10.1109/TGRS.2012.2227333}}


\bibitem[Lei and Saeed(2024)]%
        {lei2024do}
\bibfield{author}{\bibinfo{person}{Demi Lei} {and} \bibinfo{person}{Ahmed
  Saeed}.} \bibinfo{year}{2024}\natexlab{}.
\newblock \showarticletitle{Do We Need a Million Satellites in Orbit?
  Constellation-as-a-Service with Modular Satellites: Challenges and
  Opportunities}. In \bibinfo{booktitle}{\emph{Proceedings of the 2nd
  International Workshop on LEO Networking and Communication}} (Washington, DC,
  USA) \emph{(\bibinfo{series}{LEO-NET '24})}. \bibinfo{publisher}{Association
  for Computing Machinery}, \bibinfo{address}{New York, NY, USA},
  \bibinfo{pages}{61--66}.
\newblock
\href{https://doi.org/10.1145/3697253.3697262}{doi:\nolinkurl{10.1145/3697253.3697262}}


\bibitem[Leyva-Mayorga et~al\mbox{.}(2023)]%
        {leyva2023satellite}
\bibfield{author}{\bibinfo{person}{Israel Leyva-Mayorga}, \bibinfo{person}{Marc
  Martinez-Gost}, \bibinfo{person}{Marco Moretti}, \bibinfo{person}{Ana
  P{\'e}rez-Neira}, \bibinfo{person}{Miguel~{\'A}ngel V{\'a}zquez},
  \bibinfo{person}{Petar Popovski}, {and} \bibinfo{person}{Beatriz Soret}.}
  \bibinfo{year}{2023}\natexlab{}.
\newblock \showarticletitle{Satellite Edge Computing for Real-Time and
  Very-High Resolution Earth Observation}.
\newblock \bibinfo{journal}{\emph{IEEE Transactions on Communications}}
  \bibinfo{volume}{71}, \bibinfo{number}{10} (\bibinfo{date}{July}
  \bibinfo{year}{2023}), \bibinfo{pages}{6180--6194}.
\newblock
\showISSN{0090-6778}
\href{https://doi.org/10.1109/TCOMM.2023.3296584}{doi:\nolinkurl{10.1109/TCOMM.2023.3296584}}


\bibitem[Lu et~al\mbox{.}(2024)]%
        {lu2024onboard}
\bibfield{author}{\bibinfo{person}{Sha Lu}, \bibinfo{person}{Eriita Jones},
  \bibinfo{person}{Liang Zhao}, \bibinfo{person}{Yu Sun}, \bibinfo{person}{Kai
  Qin}, \bibinfo{person}{Jixue Liu}, \bibinfo{person}{Jiuyong Li},
  \bibinfo{person}{Prabath Abeysekara}, \bibinfo{person}{Norman Mueller},
  \bibinfo{person}{Simon Oliver}, \bibinfo{person}{Jim O'Hehir}, {and}
  \bibinfo{person}{Stefan Peters}.} \bibinfo{year}{2024}\natexlab{}.
\newblock \showarticletitle{Onboard AI for Fire Smoke Detection using
  Hyperspectral Imagery: an Emulation for the Upcoming Kanyini Hyperscout-2
  Mission}.
\newblock \bibinfo{journal}{\emph{IEEE Journal of Selected Topics in Applied
  Earth Observations and Remote Sensing}}  \bibinfo{volume}{17}
  (\bibinfo{date}{April} \bibinfo{year}{2024}), \bibinfo{pages}{9629--9640}.
\newblock
\showISSN{1939-1404}
\href{https://doi.org/10.1109/JSTARS.2024.3394574}{doi:\nolinkurl{10.1109/JSTARS.2024.3394574}}


\bibitem[Manning and Bowman(2023)]%
        {nasaglossary}
\bibfield{author}{\bibinfo{person}{Catherine~G. Manning} {and}
  \bibinfo{person}{Abigail Bowman}.} \bibinfo{year}{2023}\natexlab{}.
\newblock \bibinfo{booktitle}{\emph{SCaN Glossary}}.
\newblock National Aeronautics and Space Administration, Space Communications
  and Navigation (SCaN) Program.
\newblock
\urldef\tempurl%
\url{https://www.nasa.gov/reference/scan-glossary/}
\showURL{%
Retrieved August 28, 2024 from \tempurl}


\bibitem[Manor(2018)]%
        {gvisor-gcf}
\bibfield{author}{\bibinfo{person}{Eyal Manor}.}
  \bibinfo{year}{2018}\natexlab{}.
\newblock \bibinfo{booktitle}{\emph{Bringing the best of serverless to you}}.
\newblock Google Cloud Platform.
\newblock
\urldef\tempurl%
\url{https://cloudplatform.googleblog.com/2018/07/bringing-the-best-of-serverless-to-you.html}
\showURL{%
Retrieved February 12, 2024 from \tempurl}


\bibitem[McDowell(2020)]%
        {mcdowell2020low}
\bibfield{author}{\bibinfo{person}{Jonathan~C. McDowell}.}
  \bibinfo{year}{2020}\natexlab{}.
\newblock \showarticletitle{The Low Earth Orbit Satellite Population and
  Impacts of the SpaceX Starlink Constellation}.
\newblock \bibinfo{journal}{\emph{The Astrophysical Journal Letters}}
  \bibinfo{volume}{892}, \bibinfo{number}{2}, Article \bibinfo{articleno}{L36}
  (\bibinfo{date}{April} \bibinfo{year}{2020}), \bibinfo{numpages}{10}~pages.
\newblock
\href{https://doi.org/10.3847/2041-8213/ab8016}{doi:\nolinkurl{10.3847/2041-8213/ab8016}}


\bibitem[Moebius et~al\mbox{.}(2024)]%
        {moebius2024unikernel}
\bibfield{author}{\bibinfo{person}{Felix Moebius}, \bibinfo{person}{Tobias
  Pfandzelter}, {and} \bibinfo{person}{David Bermbach}.}
  \bibinfo{year}{2024}\natexlab{}.
\newblock \showarticletitle{Are Unikernels Ready for Serverless on the Edge?}.
  In \bibinfo{booktitle}{\emph{Proceedings of the 12th IEEE International
  Conference on Cloud Engineering}} (Paphos, Cyprus)
  \emph{(\bibinfo{series}{IC2E '24})}. \bibinfo{publisher}{IEEE},
  \bibinfo{address}{New York, NY, USA}, \bibinfo{pages}{133--143}.
\newblock
\href{https://doi.org/10.1109/IC2E61754.2024.00022}{doi:\nolinkurl{10.1109/IC2E61754.2024.00022}}


\bibitem[Nieto-Peroy and Emami(2019)]%
        {nieto2019cubesat}
\bibfield{author}{\bibinfo{person}{Crist{\'o}bal Nieto-Peroy} {and}
  \bibinfo{person}{M.~Reza Emami}.} \bibinfo{year}{2019}\natexlab{}.
\newblock \showarticletitle{CubeSat Mission: From Design to Operation}.
\newblock \bibinfo{journal}{\emph{Applied Sciences}} \bibinfo{volume}{9},
  \bibinfo{number}{15}, Article \bibinfo{articleno}{3110} (\bibinfo{date}{Aug.}
  \bibinfo{year}{2019}), \bibinfo{numpages}{24}~pages.
\newblock
\showISSN{2076-3417}
\href{https://doi.org/10.3390/app9153110}{doi:\nolinkurl{10.3390/app9153110}}


\bibitem[O'Donnell et~al\mbox{.}(2023)]%
        {o2023extension}
\bibfield{author}{\bibinfo{person}{Kathryn O'Donnell}, \bibinfo{person}{Meghan
  Weber}, \bibinfo{person}{Joy Fasnacht}, \bibinfo{person}{Jeff Maynard},
  \bibinfo{person}{Margaret Cote}, {and} \bibinfo{person}{Shayn Hawthorne}.}
  \bibinfo{year}{2023}\natexlab{}.
\newblock \showarticletitle{Extension of cloud computing to small satellites}.
  In \bibinfo{booktitle}{\emph{Proceedings of the 37th Annual Small Satellite
  Conference}} (Logan, UT, USA) \emph{(\bibinfo{series}{SmallSat '23})}.
  \bibinfo{publisher}{Utah State University}, \bibinfo{address}{Logan, UT,
  USA}, \bibinfo{pages}{304--318}.
\newblock


\bibitem[Pemberton and Schleier-Smith(2019)]%
        {pemberton2019serverless}
\bibfield{author}{\bibinfo{person}{Nathan Pemberton} {and}
  \bibinfo{person}{Johann Schleier-Smith}.} \bibinfo{year}{2019}\natexlab{}.
\newblock \showarticletitle{The serverless data center: Hardware disaggregation
  meets serverless computing}. In \bibinfo{booktitle}{\emph{Proceedings of the
  The First Workshop on Resource Disaggregation}} (Providence, RI, USA)
  \emph{(\bibinfo{series}{WORD '19})}.
\newblock
\urldef\tempurl%
\url{http://word19.ece.cornell.edu/serverless_data_center.pdf}
\showURL{%
\tempurl}


\bibitem[Pfandzelter and Bermbach(2020)]%
        {pfandzelter2020tinyfaas}
\bibfield{author}{\bibinfo{person}{Tobias Pfandzelter} {and}
  \bibinfo{person}{David Bermbach}.} \bibinfo{year}{2020}\natexlab{}.
\newblock \showarticletitle{tinyFaaS: A Lightweight FaaS Platform for Edge
  Environments}. In \bibinfo{booktitle}{\emph{Proceedings of the Second IEEE
  International Conference on Fog Computing}} (Sydney, NSW, Australia)
  \emph{(\bibinfo{series}{ICFC 2020})}. \bibinfo{publisher}{IEEE},
  \bibinfo{address}{New York, NY, USA}, \bibinfo{pages}{17--24}.
\newblock
\href{https://doi.org/10.1109/ICFC49376.2020.00011}{doi:\nolinkurl{10.1109/ICFC49376.2020.00011}}


\bibitem[Pfandzelter and Bermbach(2023)]%
        {pfandzelter2023failure}
\bibfield{author}{\bibinfo{person}{Tobias Pfandzelter} {and}
  \bibinfo{person}{David Bermbach}.} \bibinfo{year}{2023}\natexlab{}.
\newblock \showarticletitle{Edge Computing in Low-Earth Orbit -- What Could
  Possibly Go Wrong?}. In \bibinfo{booktitle}{\emph{Proceedings of the the 1st
  ACM Workshop on LEO Networking and Communication 2023}} (Madrid, Spain)
  \emph{(\bibinfo{series}{LEO-NET '23})}. \bibinfo{publisher}{Association for
  Computing Machinery}, \bibinfo{address}{New York, NY, USA},
  \bibinfo{pages}{19--24}.
\newblock
\href{https://doi.org/10.1145/3614204.3616106}{doi:\nolinkurl{10.1145/3614204.3616106}}


\bibitem[Pfandzelter and Bermbach(2024)]%
        {pfandzelter2024komet}
\bibfield{author}{\bibinfo{person}{Tobias Pfandzelter} {and}
  \bibinfo{person}{David Bermbach}.} \bibinfo{year}{2024}\natexlab{}.
\newblock \showarticletitle{Komet: A Serverless Platform for Low-Earth Orbit
  Edge Services}. In \bibinfo{booktitle}{\emph{Proceedings of the 15th ACM
  Symposium on Cloud Computing}} (Redmond, WA, USA)
  \emph{(\bibinfo{series}{SoCC '24})}. \bibinfo{publisher}{Association for
  Computing Machinery}, \bibinfo{address}{New York, NY, USA},
  \bibinfo{pages}{866--882}.
\newblock
\href{https://doi.org/10.1145/3698038.3698517}{doi:\nolinkurl{10.1145/3698038.3698517}}


\bibitem[Pfandzelter et~al\mbox{.}(2021)]%
        {pfandzelter2021towards}
\bibfield{author}{\bibinfo{person}{Tobias Pfandzelter},
  \bibinfo{person}{Jonathan Hasenburg}, {and} \bibinfo{person}{David
  Bermbach}.} \bibinfo{year}{2021}\natexlab{}.
\newblock \showarticletitle{Towards a Computing Platform for the LEO Edge}. In
  \bibinfo{booktitle}{\emph{Proceedings of the 4th International Workshop on
  Edge Systems, Analytics and Networking}} (Online, United Kingdom)
  \emph{(\bibinfo{series}{EdgeSys '21})}. \bibinfo{publisher}{Association for
  Computing Machinery}, \bibinfo{address}{New York, NY, USA},
  \bibinfo{pages}{43--48}.
\newblock
\href{https://doi.org/10.1145/3434770.3459736}{doi:\nolinkurl{10.1145/3434770.3459736}}


\bibitem[Raffaelle(2023)]%
        {raffaele2023cubesat}
\bibfield{author}{\bibinfo{person}{Ryne~P. Raffaelle}.}
  \bibinfo{year}{2023}\natexlab{}.
\newblock \showarticletitle{9 - Introduction to CubeSat power systems}.
\newblock In \bibinfo{booktitle}{\emph{Next Generation CubeSats and
  SmallSats}}, \bibfield{editor}{\bibinfo{person}{Francesco Branz},
  \bibinfo{person}{Chantal Cappelletti}, \bibinfo{person}{Antonio~J. Ricco},
  {and} \bibinfo{person}{John~W. Hines}} (Eds.). \bibinfo{publisher}{Elsevier},
  \bibinfo{pages}{201--221}.
\newblock
\showISBNx{978-0-12-824541-5}
\href{https://doi.org/10.1016/B978-0-12-824541-5.00008-X}{doi:\nolinkurl{10.1016/B978-0-12-824541-5.00008-X}}


\bibitem[Raith et~al\mbox{.}(2023)]%
        {raith2023serverless}
\bibfield{author}{\bibinfo{person}{Philipp Raith}, \bibinfo{person}{Stefan
  Nastic}, {and} \bibinfo{person}{Schahram Dustdar}.}
  \bibinfo{year}{2023}\natexlab{}.
\newblock \showarticletitle{Serverless Edge Computing -- Where We Are and What
  Lies Ahead}.
\newblock \bibinfo{journal}{\emph{IEEE Internet Computing}}
  \bibinfo{volume}{27}, \bibinfo{number}{3} (\bibinfo{date}{May}
  \bibinfo{year}{2023}), \bibinfo{pages}{50--64}.
\newblock
\showISSN{1089-7801}
\href{https://doi.org/10.1109/MIC.2023.3260939}{doi:\nolinkurl{10.1109/MIC.2023.3260939}}


\bibitem[Ronneberger et~al\mbox{.}(2015)]%
        {ronneberger2015u}
\bibfield{author}{\bibinfo{person}{Olaf Ronneberger}, \bibinfo{person}{Philipp
  Fischer}, {and} \bibinfo{person}{Thomas Brox}.}
  \bibinfo{year}{2015}\natexlab{}.
\newblock \showarticletitle{U-Net: Convolutional Networks for Biomedical Image
  Segmentation}. In \bibinfo{booktitle}{\emph{Proceedings of the Medical image
  computing and computer-assisted intervention}} (Munich, Germany)
  \emph{(\bibinfo{series}{MICCAI '15})}. \bibinfo{publisher}{Springer},
  \bibinfo{address}{Cham, Switzerland}, \bibinfo{pages}{234--241}.
\newblock
\href{https://doi.org/10.1007/978-3-319-24574-4_28}{doi:\nolinkurl{10.1007/978-3-319-24574-4_28}}


\bibitem[Ryan et~al\mbox{.}(2022)]%
        {ryan2022impact}
\bibfield{author}{\bibinfo{person}{Robert~G. Ryan}, \bibinfo{person}{Eloise~A.
  Marais}, \bibinfo{person}{Chloe~J. Balhatchet}, {and}
  \bibinfo{person}{Sebastian~D. Eastham}.} \bibinfo{year}{2022}\natexlab{}.
\newblock \showarticletitle{Impact of Rocket Launch and Space Debris Air
  Pollutant Emissions on Stratospheric Ozone and Global Climate}.
\newblock \bibinfo{journal}{\emph{Earth's Future}} \bibinfo{volume}{10},
  \bibinfo{number}{6}, Article \bibinfo{articleno}{e2021EF002612}
  (\bibinfo{date}{June} \bibinfo{year}{2022}).
\newblock
\showISSN{2328-4277}
\href{https://doi.org/10.1029/2021EF002612}{doi:\nolinkurl{10.1029/2021EF002612}}


\bibitem[Sahraei et~al\mbox{.}(2023)]%
        {sahraei2023xfaas}
\bibfield{author}{\bibinfo{person}{Alireza Sahraei}, \bibinfo{person}{Soteris
  Demetriou}, \bibinfo{person}{Amirali Sobhgol}, \bibinfo{person}{Haoran
  Zhang}, \bibinfo{person}{Abhigna Nagaraja}, \bibinfo{person}{Neeraj Pathak},
  \bibinfo{person}{Girish Joshi}, \bibinfo{person}{Carla Souza},
  \bibinfo{person}{Bo Huang}, \bibinfo{person}{Wyatt Cook},
  \bibinfo{person}{Andrii Golovei}, \bibinfo{person}{Pradeep Venkat},
  \bibinfo{person}{Andrew Mcfague}, \bibinfo{person}{Dimitrios Skarlatos},
  \bibinfo{person}{Vipul Patel}, \bibinfo{person}{Ravinder Thind},
  \bibinfo{person}{Ernesto Gonzalez}, \bibinfo{person}{Yun Jin}, {and}
  \bibinfo{person}{Chunqiang Tang}.} \bibinfo{year}{2023}\natexlab{}.
\newblock \showarticletitle{XFaaS: Hyperscale and Low Cost Serverless Functions
  at Meta}. In \bibinfo{booktitle}{\emph{Proceedings of the 29th Symposium on
  Operating Systems Principles}} (Koblenz, Germany)
  \emph{(\bibinfo{series}{SOSP '23})}. \bibinfo{publisher}{Association for
  Computing Machinery}, \bibinfo{address}{New York, NY, USA},
  \bibinfo{pages}{231--246}.
\newblock
\href{https://doi.org/10.1145/3600006.3613155}{doi:\nolinkurl{10.1145/3600006.3613155}}


\bibitem[Samer~Alashhab and Gil(2019)]%
        {alashhab2019precise}
\bibfield{author}{\bibinfo{person}{Antonio~Pertusa Samer~Alashhab,
  Antonio-Javier~Gallego} {and} \bibinfo{person}{Pablo Gil}.}
  \bibinfo{year}{2019}\natexlab{}.
\newblock \showarticletitle{Precise Ship Location With CNN Filter Selection
  From Optical Aerial Images}.
\newblock \bibinfo{journal}{\emph{IEEE Access}}  \bibinfo{volume}{7}
  (\bibinfo{date}{July} \bibinfo{year}{2019}), \bibinfo{pages}{96567--96582}.
\newblock
\showISSN{2169--3536}
\href{https://doi.org/10.1109/ACCESS.2019.2929080}{doi:\nolinkurl{10.1109/ACCESS.2019.2929080}}


\bibitem[Schirmer et~al\mbox{.}(2023)]%
        {schirmer2023profaastinate}
\bibfield{author}{\bibinfo{person}{Trever Schirmer}, \bibinfo{person}{Valentin
  Carl}, \bibinfo{person}{Tobias Pfandzelter}, {and} \bibinfo{person}{David
  Bermbach}.} \bibinfo{year}{2023}\natexlab{}.
\newblock \showarticletitle{ProFaaStinate: Delaying Serverless Function Calls
  to Optimize Platform Performance}. In \bibinfo{booktitle}{\emph{Proceedings
  of the 9th International Workshop on Serverless Computing}} (Bologna, Italy)
  \emph{(\bibinfo{series}{WoSC '23})}. \bibinfo{publisher}{Association for
  Computing Machinery}, \bibinfo{address}{New York, NY, USA},
  \bibinfo{pages}{1--6}.
\newblock
\href{https://doi.org/10.1145/3631295.3631393}{doi:\nolinkurl{10.1145/3631295.3631393}}


\bibitem[Shamendra et~al\mbox{.}(2023)]%
        {shamendra2023trufaas}
\bibfield{author}{\bibinfo{person}{Avishka Shamendra}, \bibinfo{person}{Binoy
  Peries}, \bibinfo{person}{Gayangi Seneviratne}, {and}
  \bibinfo{person}{Sunimal Rathnayake}.} \bibinfo{year}{2023}\natexlab{}.
\newblock \showarticletitle{TruFaaS-Trust Verification Framework for FaaS}. In
  \bibinfo{booktitle}{\emph{Proceedings of the International Conference on
  Ubiquitous Security}} (Exeter, UK) \emph{(\bibinfo{series}{UbiSec '23})}.
  \bibinfo{publisher}{Springer}, \bibinfo{address}{Cham, Switzerland},
  \bibinfo{pages}{304--318}.
\newblock
\href{https://doi.org/10.1007/978-981-97-1274-8_20}{doi:\nolinkurl{10.1007/978-981-97-1274-8_20}}


\bibitem[Shi and Dustdar(2016)]%
        {shi2016edge}
\bibfield{author}{\bibinfo{person}{Weisong Shi} {and} \bibinfo{person}{Schahram
  Dustdar}.} \bibinfo{year}{2016}\natexlab{}.
\newblock \showarticletitle{The Promise of Edge Computing}.
\newblock \bibinfo{journal}{\emph{Computer}} \bibinfo{volume}{49},
  \bibinfo{number}{5} (\bibinfo{date}{May} \bibinfo{year}{2016}),
  \bibinfo{pages}{78--81}.
\newblock
\showISSN{0018-9162}
\href{https://doi.org/10.1109/MC.2016.145}{doi:\nolinkurl{10.1109/MC.2016.145}}


\bibitem[Slater et~al\mbox{.}(2020)]%
        {slater2020total}
\bibfield{author}{\bibinfo{person}{Windy~S. Slater}, \bibinfo{person}{Nayana~P.
  Tiwari}, \bibinfo{person}{Tyler~M. Lovelly}, {and} \bibinfo{person}{Jesse~K.
  Mee}.} \bibinfo{year}{2020}\natexlab{}.
\newblock \showarticletitle{Total Ionizing Dose Radiation Testing of NVIDIA
  Jetson Nano GPUs}. In \bibinfo{booktitle}{\emph{Proceedings of the 2020 IEEE
  High Performance Extreme Computing Conference}} (Waltham, MA, USA)
  \emph{(\bibinfo{series}{HPEC '20})}. \bibinfo{publisher}{IEEE},
  \bibinfo{address}{New York, NY, USA}, \bibinfo{pages}{1--3}.
\newblock
\href{https://doi.org/10.1109/HPEC43674.2020.9286222}{doi:\nolinkurl{10.1109/HPEC43674.2020.9286222}}


\bibitem[Soret et~al\mbox{.}(2024)]%
        {soret2024semantic}
\bibfield{author}{\bibinfo{person}{Beatriz Soret}, \bibinfo{person}{Israel
  Leyva-Mayorga}, \bibinfo{person}{Antonio~M. Mercado-Martínez},
  \bibinfo{person}{Marco Moretti}, \bibinfo{person}{Antonio Jurado-Navas},
  \bibinfo{person}{Marc Martinez-Gost}, \bibinfo{person}{Celia S\'anchez~de
  Miguel}, \bibinfo{person}{Ainoa Salas-Prendes}, {and} \bibinfo{person}{Petar
  Popovski}.} \bibinfo{year}{2024}\natexlab{}.
\newblock \showarticletitle{Semantic and goal-oriented edge computing for
  satellite Earth Observation}.
\newblock  (\bibinfo{date}{Aug.} \bibinfo{year}{2024}).
\newblock
\showeprint{2408.15639}


\bibitem[Speed(2021)]%
        {speed2021hpe}
\bibfield{author}{\bibinfo{person}{Richard Speed}.}
  \bibinfo{year}{2021}\natexlab{}.
\newblock \bibinfo{booktitle}{\emph{HPE Spaceborne Computer-2 slips off the
  shelf -- and off the planet: Boxen heading to ISS}}.
\newblock The Register.
\newblock
\urldef\tempurl%
\url{https://www.theregister.com/2021/02/11/hpe_spaceborne_2_iss/}
\showURL{%
Retrieved April 9, 2025 from \tempurl}


\bibitem[Sumbul et~al\mbox{.}(2019)]%
        {sumbul2019bigearthnet}
\bibfield{author}{\bibinfo{person}{Gencer Sumbul}, \bibinfo{person}{Marcela
  Charfuelan}, \bibinfo{person}{Beg{\"u}m Demir}, {and} \bibinfo{person}{Volker
  Markl}.} \bibinfo{year}{2019}\natexlab{}.
\newblock \showarticletitle{Bigearthnet: A Large-Scale Benchmark Archive for
  Remote Sensing Image Understanding}. In \bibinfo{booktitle}{\emph{Proceedings
  of the 2019 IEEE International Geoscience and Remote Sensing Symposium}}
  (Yokohama, Japan) \emph{(\bibinfo{series}{IGARSS '19})}.
  \bibinfo{publisher}{IEEE}, \bibinfo{address}{New York, NY, USA},
  \bibinfo{pages}{5901--5904}.
\newblock
\href{https://doi.org/10.1109/IGARSS.2019.8900532}{doi:\nolinkurl{10.1109/IGARSS.2019.8900532}}


\bibitem[Varon et~al\mbox{.}(2021)]%
        {varon2021high}
\bibfield{author}{\bibinfo{person}{Daniel~J Varon}, \bibinfo{person}{Dylan
  Jervis}, \bibinfo{person}{Jason McKeever}, \bibinfo{person}{Ian Spence},
  \bibinfo{person}{David Gains}, {and} \bibinfo{person}{Daniel~J Jacob}.}
  \bibinfo{year}{2021}\natexlab{}.
\newblock \showarticletitle{High-frequency monitoring of anomalous methane
  point sources with multispectral Sentinel-2 satellite observations}.
\newblock \bibinfo{journal}{\emph{Atmospheric Measurement Techniques}}
  \bibinfo{volume}{14}, \bibinfo{number}{4} (\bibinfo{date}{April}
  \bibinfo{year}{2021}), \bibinfo{pages}{2771--2785}.
\newblock
\showISSN{1867--8548}
\href{https://doi.org/10.5194/amt-14-2771-2021}{doi:\nolinkurl{10.5194/amt-14-2771-2021}}


\bibitem[Vasisht et~al\mbox{.}(2021)]%
        {vasisht2021l2d2}
\bibfield{author}{\bibinfo{person}{Deepak Vasisht}, \bibinfo{person}{Jayanth
  Shenoy}, {and} \bibinfo{person}{Ranveer Chandra}.}
  \bibinfo{year}{2021}\natexlab{}.
\newblock \showarticletitle{L2D2: low latency distributed downlink for LEO
  satellites}. In \bibinfo{booktitle}{\emph{Proceedings of the 2021 ACM SIGCOMM
  2021 Conference}} (Virtual Event, USA) \emph{(\bibinfo{series}{SIGCOMM
  '21})}. \bibinfo{publisher}{Association for Computing Machinery},
  \bibinfo{address}{New York, NY, USA}, \bibinfo{pages}{151--164}.
\newblock
\href{https://doi.org/10.1145/3452296.3472932}{doi:\nolinkurl{10.1145/3452296.3472932}}


\bibitem[Villas~Boas et~al\mbox{.}(2023)]%
        {villasboas2023innovative}
\bibfield{author}{\bibinfo{person}{Danton José~Fortes Villas~Boas},
  \bibinfo{person}{José~Bezerra Pessoa~Filho}, \bibinfo{person}{Alison de
  Oliveira~Moraes}, {and} \bibinfo{person}{Carlos~Henrique Melo~Souza}.}
  \bibinfo{year}{2023}\natexlab{}.
\newblock \showarticletitle{17 - Innovative and low-cost launch systems}.
\newblock In \bibinfo{booktitle}{\emph{Next Generation CubeSats and
  SmallSats}}, \bibfield{editor}{\bibinfo{person}{Francesco Branz},
  \bibinfo{person}{Chantal Cappelletti}, \bibinfo{person}{Antonio~J. Ricco},
  {and} \bibinfo{person}{John~W. Hines}} (Eds.). \bibinfo{publisher}{Elsevier},
  \bibinfo{pages}{403--419}.
\newblock
\showISBNx{978-0-12-824541-5}
\href{https://doi.org/10.1016/B978-0-12-824541-5.00005-4}{doi:\nolinkurl{10.1016/B978-0-12-824541-5.00005-4}}


\bibitem[Wang et~al\mbox{.}(2023)]%
        {wang2023satellite}
\bibfield{author}{\bibinfo{person}{Chao Wang}, \bibinfo{person}{Yiran Zhang},
  \bibinfo{person}{Qing Li}, \bibinfo{person}{Ao Zhou}, {and}
  \bibinfo{person}{Shangguang Wang}.} \bibinfo{year}{2023}\natexlab{}.
\newblock \showarticletitle{Satellite Computing: A Case Study of Cloud-Native
  Satellites}. In \bibinfo{booktitle}{\emph{Proceedings of the 2023 IEEE
  International Conference on Edge Computing and Communications}} (Chicago, IL,
  USA) \emph{(\bibinfo{series}{EDGE '23})}. \bibinfo{publisher}{IEEE},
  \bibinfo{address}{New York, NY, USA}, \bibinfo{pages}{262--270}.
\newblock
\href{https://doi.org/10.1109/EDGE60047.2023.00048}{doi:\nolinkurl{10.1109/EDGE60047.2023.00048}}


\bibitem[Wang and Li(2023)]%
        {wang2023vision}
\bibfield{author}{\bibinfo{person}{Shangguang Wang} {and} \bibinfo{person}{Qing
  Li}.} \bibinfo{year}{2023}\natexlab{}.
\newblock \showarticletitle{Satellite Computing: Vision and Challenges}.
\newblock \bibinfo{journal}{\emph{IEEE Internet of Things Journal}}
  \bibinfo{volume}{10}, \bibinfo{number}{24} (\bibinfo{date}{Aug.}
  \bibinfo{year}{2023}), \bibinfo{pages}{22514--22529}.
\newblock
\showISSN{2327-4662}
\href{https://doi.org/10.1109/JIOT.2023.3303346}{doi:\nolinkurl{10.1109/JIOT.2023.3303346}}


\bibitem[Wang et~al\mbox{.}(2021)]%
        {wang2021tiansuan}
\bibfield{author}{\bibinfo{person}{Shangguang Wang}, \bibinfo{person}{Qing Li},
  \bibinfo{person}{Mengwei Xu}, \bibinfo{person}{Xiao Ma}, \bibinfo{person}{Ao
  Zhou}, {and} \bibinfo{person}{Qibo Sun}.} \bibinfo{year}{2021}\natexlab{}.
\newblock \showarticletitle{Tiansuan Constellation: An Open Research Platform}.
  In \bibinfo{booktitle}{\emph{Proceedings of the 2021 IEEE International
  Conference on Edge Computing}} (Chicago, IL, USA)
  \emph{(\bibinfo{series}{EDGE '21})}. \bibinfo{publisher}{IEEE},
  \bibinfo{address}{New York, NY, USA}, \bibinfo{pages}{94--101}.
\newblock
\href{https://doi.org/10.1109/EDGE53862.2021.00022}{doi:\nolinkurl{10.1109/EDGE53862.2021.00022}}


\bibitem[Xie et~al\mbox{.}(2021)]%
        {xie2021serverless}
\bibfield{author}{\bibinfo{person}{Renchao Xie}, \bibinfo{person}{Qinqin Tang},
  \bibinfo{person}{Shi Qiao}, \bibinfo{person}{Han Zhu},
  \bibinfo{person}{F.~Richard Yu}, {and} \bibinfo{person}{Tao Huang}.}
  \bibinfo{year}{2021}\natexlab{}.
\newblock \showarticletitle{When Serverless Computing Meets Edge Computing:
  Architecture, Challenges, and Open Issuess}.
\newblock \bibinfo{journal}{\emph{IEEE Wireless Communications}}
  \bibinfo{volume}{28}, \bibinfo{number}{5} (\bibinfo{date}{July}
  \bibinfo{year}{2021}), \bibinfo{pages}{126--133}.
\newblock
\showISSN{1536-1284}
\href{https://doi.org/10.1109/MWC.001.2000466}{doi:\nolinkurl{10.1109/MWC.001.2000466}}


\bibitem[Xing et~al\mbox{.}(2024)]%
        {xing2024deciphering}
\bibfield{author}{\bibinfo{person}{Ruolin Xing}, \bibinfo{person}{Mengwei Xu},
  \bibinfo{person}{Ao Zhou}, \bibinfo{person}{Qing Li}, \bibinfo{person}{Yiran
  Zhang}, \bibinfo{person}{Feng Qian}, {and} \bibinfo{person}{Shangguang
  Wang}.} \bibinfo{year}{2024}\natexlab{}.
\newblock \showarticletitle{Deciphering the Enigma of Satellite Computing with
  COTS Devices: Measurement and Analysis}. In
  \bibinfo{booktitle}{\emph{Proceedings of the 30th Annual International
  Conference on Mobile Computing and Networking}} (Washington D.C., DC, USA)
  \emph{(\bibinfo{series}{MobiCom '24})}. \bibinfo{publisher}{Association for
  Computing Machinery}, \bibinfo{address}{New York, NY, USA},
  \bibinfo{pages}{420--435}.
\newblock
\href{https://doi.org/10.1145/3636534.3649371}{doi:\nolinkurl{10.1145/3636534.3649371}}


\bibitem[Yendler(2021)]%
        {yendler2021thermal}
\bibfield{author}{\bibinfo{person}{Boris Yendler}.}
  \bibinfo{year}{2021}\natexlab{}.
\newblock \showarticletitle{16 - Thermal control system}.
\newblock In \bibinfo{booktitle}{\emph{Cubesat Handbook}},
  \bibfield{editor}{\bibinfo{person}{Chantal Cappelletti},
  \bibinfo{person}{Simone Battistini}, {and} \bibinfo{person}{Benjamin~K.
  Malphrus}} (Eds.). \bibinfo{publisher}{Academic Press},
  \bibinfo{pages}{303--317}.
\newblock
\showISBNx{978-0-12-817884-3}
\href{https://doi.org/10.1016/B978-0-12-817884-3.00016-3}{doi:\nolinkurl{10.1016/B978-0-12-817884-3.00016-3}}


\bibitem[Zengshan et~al\mbox{.}(2024)]%
        {yin2024comprehensive}
\bibfield{author}{\bibinfo{person}{Yin Zengshan}, \bibinfo{person}{Wu
  Changhao}, \bibinfo{person}{Guo Chongbin}, \bibinfo{person}{Li Yuanchun},
  \bibinfo{person}{Xu Mengwei}, \bibinfo{person}{Gao Weiwei}, {and}
  \bibinfo{person}{Chi Chuanxiu}.} \bibinfo{year}{2024}\natexlab{}.
\newblock \showarticletitle{A comprehensive survey of orbital edge computing:
  Systems, applications, and algorithms}.
\newblock \bibinfo{journal}{\emph{Chinese Journal of Aeronautics}}, Article
  \bibinfo{articleno}{3316} (\bibinfo{date}{Nov.} \bibinfo{year}{2024}),
  \bibinfo{numpages}{30}~pages.
\newblock
\showISSN{1000-9361}
\href{https://doi.org/10.1016/j.cja.2024.11.026}{doi:\nolinkurl{10.1016/j.cja.2024.11.026}}
\showeprint{2306.00275}


\bibitem[Zhang et~al\mbox{.}(2022)]%
        {zhang2022lightweight}
\bibfield{author}{\bibinfo{person}{Denghui Zhang}, \bibinfo{person}{Lijing
  Ren}, \bibinfo{person}{Muhammad Shafiq}, {and} \bibinfo{person}{Zhaoquan
  Gu}.} \bibinfo{year}{2022}\natexlab{}.
\newblock \showarticletitle{A Lightweight Privacy-Preserving System for the
  Security of Remote Sensing Images on IoT}.
\newblock \bibinfo{journal}{\emph{Remote Sensing}} \bibinfo{volume}{14},
  \bibinfo{number}{24}, Article \bibinfo{articleno}{6371} (\bibinfo{date}{Dec.}
  \bibinfo{year}{2022}), \bibinfo{numpages}{16}~pages.
\newblock
\showISSN{2072-4292}
\href{https://doi.org/10.3390/rs14246371}{doi:\nolinkurl{10.3390/rs14246371}}


\end{thebibliography}

\end{document}